\documentclass[aps,pra,twocolumn,groupedaddress,showpacs]{revtex4-1}
\usepackage{hyperref}
\usepackage[T1]{fontenc}
\usepackage[utf8]{inputenc}
\usepackage{amsmath}
\usepackage{amssymb}
\usepackage{esint}
\usepackage{color}
\usepackage{graphicx}
\usepackage{epsfig}
\usepackage{pstricks}           % e.g. colors can be used in the text
\usepackage{amsfonts}
\usepackage{dcolumn}
\usepackage{bm}
\usepackage{overpic}
\usepackage{pict2e}
\newcommand{\etal}{\textit{et al.}~}
\graphicspath{{figures/}}

\begin{document}

\preprint{APS/123-QED}

\title{High-harmonic generation in solids}% Force line breaks with \\
%\thanks{}%
\author{Francisco Navarrete$^1$, Marcelo F. Ciappina$^2$, and Uwe Thumm$^1$$\footnote{Corresponding author: thumm@phys.ksu.edu}$}
\affiliation{$^1$ Department of Physics, Kansas State University,
Manhattan, Kansas 66506, USA}
\affiliation{$^2$ Institute of Physics of the ASCR, ELI-Beamlines
project, Na Slovance 2, 182 21 Prague, Czech Republic}

%\email{navarrete@phys.ksu.edu}
%\email{marcelo.ciappina@eli-beams.eu}
%\email{thumm@phys.ksu.edu}

\date{\today}% It is always \today, today,

\begin{abstract}
We analytically and numerically investigate the emission of
high-harmonic radiation from model solids by intense few-cycle
mid-infrared laser pulses. In single-active-electron approximation,
we expand the active electron's wavefunction in a basis of adiabatic
Houston states and describe the solid's electronic band structure in
terms of an adjustable Kronig-Penney model potential. For
high-harmonic generation (HHG) from MgO crystals, we examine spectra
from two-band and converged multiband numerical calculations. We
discuss the characteristics of intra- and interband contributions to
the HHG spectrum for computations including initial crystal momenta
either from the $\Gamma-$point at the center of the first Brioullin
zone (BZ) only or from the entire first BZ, demonstrating relevant
contributions from the entire first BZ. From the numerically
calculated spectra we derive cutoff harmonic orders as a function of
the laser peak intensity that compare favorably with our analytical
stationary-phase-approximation predictions and published data.
\end{abstract}

% PACS, the Physics and AstronWith this definition of the phases omy
\maketitle

%\tableofcontents

\section{\label{sec:level1} Introduction}

Exposed to intense laser fields gases and solids emit a spectrum of
radiation that is strongly enhanced at and near frequencies
corresponding to multiples of the driving laser frequency. Over the
past two decades, this high-order harmonic generation (HHG) process
has been carefully investigated in atomic gases, and the underlying
generation mechanism - the emission of radiation by
laser-electric-field-driven rescattered electrons - is well
understood~\cite{le2009pra,krausz2009rmp}. While solids are being
discussed theoretically for decades in view of their large
electronic density possibly enabling the design of high-intensity
sources of harmonic
radiation~\cite{pla1992prb,keitel1997rpp,faisal1996pra,faisal1997pra},
HHG from solids has remained a matter of
debate~\cite{ghimire2011natphys,wu2015pra,vampa2014prl}.
Experimentally, it was first carefully scrutinized less than a
decade ago by Ghimire~\etal~\cite{ghimire2011natphys}. Understanding
the mechanisms of HHG in solids is an area of emerging research
interest and part of the ongoing diversification of attosecond
science from the study of atoms and molecules to more complex
nanoparticles~\cite{jli2016pra,jli2017pra,jli2018prl} and
solids~\cite{thumm2014nphot,thumm2015chapt,ciappina2017rpp,kruchinin2018rmp}.
This extension of attosecond science holds promise for promoting the
development of novel table-top intense high-frequency radiation
sources and our understanding of the light-induced electron dynamics
in solids, a prerequisite for improved ultrafast electron-optical
switches~\cite{luu2016prb,reis2019}.

Compared to HHG in atomic gases, theoretical investigations of solid
HHG have indicated striking new effects, such as multiple
plateaus~\cite{ndaba2016nat,wu2015pra} in HHG spectra and a linear
dependence of the HHG cutoff frequency on the peak electric-field
strengths of the driving
laser~\cite{wu2015pra,vampa2015prb,osika2017prx,li2018pra}. These
characteristics have been revealed by numerically solving either the
time-dependent Schr\"odinger (TDSE) in single-active-electron (SAE)
approximation~\cite{mucke2011prb,hawkins2015pra,wu2015pra,li2018pra}
or semiconductor Bloch equations
(SBEs)~\cite{golde2008prb,schubert2014nphot,meier1994prl,vampa2014prl,jiang2017pra,jiang2018prl,li2018pra}.
SAE-TDSE-based numerical models have employed basis-set expansions
of the active electron's wavefunction using either static
\cite{korbman2013njp} or adiabatic~\cite{krieger1985prb,wu2015pra}
Bloch states.

SAE solutions of the TDSE can be expressed in terms of density
matrices for convenient comparison with the SBE approach that
introduces a phenomenological dephasing time to account for
relaxation processes~\cite{haug1998}. An advantage of working within
the SAE-TDSE framework is that the computational time for solving a
system with $n$ electronic bands scales linearly with $n$, while it
scales as ${n(n-1)/2}$ in SBE calculations~\cite{luu2016prb}. For
${n>20}$ this leads to approximately one order of magnitude
difference in computation time (cf., Ref.~\cite{wu2015pra}, where 51
bands are included for solving the TDSE within a static Bloch
basis). This is of relevance at high intensities of the driving
laser, where calculations with a large number of bands are required
to reveal the multiplateau structure of converged HHG spectra
~\cite{wu2015pra}. SAE models have successfully explained the main
features of HHG in atomic gases~\cite{le2009pra,krausz2009rmp},
which has motivated their transfer to describing HHG in
solids~\cite{wu2015pra,ndaba2016nat}.

In this work we develop a numerical model for solving the TDSE in
SAE approximation employing a basis-set expansion of the electronic
wavefunction in so-called ``Houston states'' that vary adiabatically
with the instantaneous driving-laser electromagnetic
field~\cite{wu2015pra,lindefelt2004sst,krieger1985prb}. The use of
an adiabatic basis is advantageous for gaining physical insight into
the underlying basic mechanisms for HHG in solids, since it allows
the identification of two distinct processes, intra- and an
interband emission, that operate in different spectral regions. In
calculating HHG spectra from solids, we pay attention to the crystal
momentum of the initial state and scrutinize contributions from
different crystal momenta in the entire first Brillouin zone (BZ).
In particular, we find that HHG spectra calculated by only including
the initial crystal momentum at the center of the BZ (the
$\Gamma-$point)~\cite{wu2015pra} noticeably differ from calculations
that include initial crystal momenta in the entire
BZ~\cite{vampa2014prl}. It has been recently proposed that the range
of crystal momenta used in SAE calculations should be used as an
adjustable parameter for getting converged numerical HHG
spectra~\cite{li2018pra}. In the present work we revisit this
suggestion and analytically determine, based on a stationary-phase
approximation, the range of initial crystal momenta needed for the
computation of HHG spectra at a given accuracy.

We organized this paper as follows. In Sec.~\ref{sec:theo} we
describe our theoretical framework. In particular, in
Sec.~\ref{subsec:tdse} we solve the TDSE by expanding the active
electron's wavefunction in an adiabatic basis (Sec.
\ref{subsubsec:houston}), compare our approach with a density-matrix
formulation of solid HHG (Sec.~\ref{subsubsec:densmat}), and  show
how the observables of interest in this work, intra- and interband
yields, are retrieved from our numerical results
(Sec.~\ref{subsubsec:intrainter}).
In Sec.~\ref{subsec:two_band} we discuss solid HHG for a simplified
two-band system distinuishing intra-
(Sec.~\ref{subsubsec:intra_yield}) and interband (Sec.
\ref{subsubsec:inter_yield}) yields.
In Sec.~\ref{subsec:ApproxInterbandYield} we analyze interband HHG
within a stationary-phase approximation
(Sec.~\ref{subsubsec:stat-phase-approx}), which allows us to
estimate the relevant range of crystal momenta $k$
(Sec.~\ref{subsubsec:relevant-k-range}) that need to be included
when adding HHG contributions from different $k$ in the first BZ
(Sec.~\ref{subsubsec:inter_yield_bz}).
In Sec.~\ref{sec:results} we present and discuss our numerical
results for HHG in a model MgO crystal. First, based on simplified
two-band calculations, we analyze in Sec.~\ref{subsec:res_two}
$k$-resolved spectra for a specific field strength of the
driving-laser pulse (Sec.~\ref{subsubsec:k_resolved}),
field-strength-dependent spectra (Sec.~\ref{subsubsec:field_dep}),
and cut-off harmonic orders as a function of the field strengths
(Sec.~\ref{subsubsec:cutoff}).
Next, in Sec.~\ref{subsec:multiband}, we analyze
field-strength-dependent HHG spectra
(Sec.~\ref{subsubsec:multi_field_dep}) and cutoff harmonic orders
(Sec.~\ref{subsubsec:cutoff-multi}) for calculations that are
converged in the number of included electronic bands. In several
appendices, we add details of our theoretical analysis. We use
atomic units (${q_e = m_e = \hbar = 1}$) throughout this work,
unless specified otherwise.

\section{\label{sec:theo}Theory}

\subsection{\label{subsec:tdse}Single-active-electron solution of the TDSE}

We solve the TDSE,
\begin{equation}
\left[\frac{1}{2}\left(\hat{p}+ A (t)\right)^2 +V(x)
\right]\left|\psi(t)\right>=i\frac{\partial}{\partial t}
\left|\psi(t)\right>, \label{TDSE_hou}
\end{equation}
subject to the interaction of the active electron with both, a
one-dimensional infinitely extended solid and an  infrared (IR)
external laser field ${E(t)}$. We represent the solid by a periodic
potential ${V(x)=V(x+a)}$, with lattice constant $a$, and the laser
field by a 10-cycle ``flat-top'' vector potential:
\begin{equation}
\begin{aligned}
A(t) &= -\int_{0}^t E\left(t'\right)dt' \\
&=\frac{A_0}{2T}\sin(\omega_0 t) \left \{
  \begin{tabular}{cl}
  $ t$ &\;, $0\leq t\leq 2T \;\,$  \\
  $2T$ &\;, $2T\leq t\leq 8T \;\,$ \\
  $(10T-t)$&\;, $8T\leq t\leq 10T \;\,$
  \end{tabular}
  \right. .
\end{aligned}
 \label{eq:pulseshape}
\end{equation}
${\hat{p}=-i\frac{\partial}{\partial x}}$ denotes the momentum
operator and $A_0$, $\omega_0$, and ${T=2\pi/\omega_0}$  the
external vector-potential amplitude, frequency, and period,
respectively. Since in our numerical simulation
(Sec.~\ref{sec:results}) the driving-laser wavelength is three
orders of magnitude larger than $a$ and the classical excursion
range of the active electron in the laser field, in solving
Eq.(\ref{TDSE_hou}) we can safely invoke the dipole approximation,
${A(x,\,t)\approx A(t)}$.

We model ${V(x)}$ as a Kronig-Penney
potential~\cite{kronig1931prs,atkins2005}, which yields the
dispersion relation
\begin{equation}
\cos (a k)= \cos (a \sqrt{2\varepsilon_{nk}} )+
\frac{V_0}{\sqrt{2\varepsilon_{nk}}} \sin (a
\sqrt{2\varepsilon_{nk}} ) \label{eq:kp}
\end{equation}
for the valence (${n=v}$) and conduction bands (${n=c}$). The
potential strength $V_0$ is adjusted to match the electronic band
structure of the solid. While Eq.~(\ref{eq:kp}) needs to be solved
numerically, the Kronig-Penney model potential provides a convenient
basis set of 2-fold-degenerate orthonormal eigenstates and allows us
to calculate transition matrix elements in a closed analytical form
(see Sec.~\ref{subsubsec:houston} below and Appendix~\ref{app_KP}).

\subsubsection{\label{subsubsec:houston}Expansion in Houston states}

Expanding solutions of Eq.~(\ref{TDSE_hou}),
\begin{equation}
\left| \psi_k (t)\right>=e^{-i  A(t)x}\sum_n B_{nk}(t)e^{\left[-i
\int_{0}^t \varepsilon_{n\kappa(t')}dt'\right]} \left|
\phi_{n\kappa(t)} \right>, \label{eq:wavefu}
\end{equation}
in terms of Houston states ${\left|\phi_{n\kappa(t)} \right>}$,
results in the set of coupled equations
\begin{equation}
\begin{aligned}
i\dot{B}_{nk}(t)&=-\sum_{n'\neq n} B_{n'k}(t) E(t)D^{nn'}_{\kappa(t)}  \\
&\times  \exp\left[i  \int_{0}^t\Delta\varepsilon_{\kappa(t)}^{nn'}
dt'\right]  \,.
\end{aligned}
\label{houstonx_c}
\end{equation}
In Eq.~(\ref{houstonx_c}) we define the energy difference between
Houston states~\cite{krieger1985prb,wu2015pra} (also referred to as
``band-gap energy'') ${\Delta\varepsilon_{\kappa(t)}^{nn'} =
\varepsilon_{n\kappa(t)}-\varepsilon_{n'\kappa(t)}}$ and  the
transition dipole moments (TDMs)
\begin{equation}
\begin{aligned}
D^{nn'}_{\kappa(t)}&= \frac{i\,P^{nn'}_{\kappa(t)}}{\Delta\varepsilon_{\kappa(t)}^{nn'}} \;\;\; (n\neq n')\\
D^{nn}_{\kappa(t)}&=0.
\end{aligned}
\label{eq:tdm}
\end{equation}
The TDMs are given in terms of the momentum-operator matrix
elements~\cite{lindefelt2004sst}
\begin{equation}
P_{\kappa(t)}^{nn'}=
\frac{1}{a}\int_0^a\phi_{n{\kappa(t)}}^*(x)\frac{1}{i}\frac{\partial}{\partial
x}\phi_{n'{\kappa(t)}}(x)\,dx . \label{eq:pnnp}
\end{equation}
The diagonal elements,
\begin{equation}
P_{\kappa(t)}^{nn}= \frac{\partial
\varepsilon_{n{\kappa(t)}}}{\partial {\kappa(t)}} \,,
\end{equation}
are related to the band energy $\varepsilon_{n{\kappa(t)}}$ and
correspond to the group velocity of an electron wave packet in band
$n$. Since the elements ${P^{nn'}_{\kappa(t)}}$ are real, the TDMs
satisfy ${D^{nn'}_{\kappa(t)}=-D^{n'n}_{\kappa(t)}}$ and
${\left(D^{nn'}_{\kappa(t)}\right)^*=D^{n'n}_{\kappa(t)}}$
(Appendix~\ref{app_KP}).

\begin{figure}[h]
\centering{}\includegraphics[width=1\linewidth]{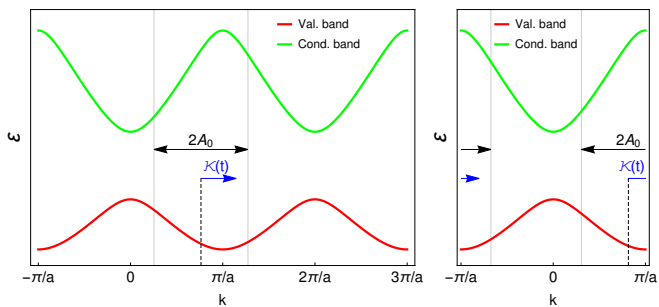}
\caption{(Color online). The two lowest dispersion curves according
Eq.~(\ref{eq:kp}) for the Kronig-Penney model in the repeated-zone
scheme (left) and in the first BZ zone (right).
The vertical black dashed line indicates a particular $k-$channel.
The single-headed  blue arrow shows the adiabatic momentum
${\kappa(t)}$ at time $t$, the double-headed arrow the maximum range
(${2A_0}$) covered by ${\kappa(t)}$ within the first BZ. }
\label{fig:fig01}
\end{figure}

Houston states are adiabatic in the field-dressed time-dependent
crystal momentum
\begin{equation}
\kappa(t)=k + A (t) \,, \label{eq:kappa_deff}
\end{equation}
and solve the Schr{\"o}dinger equation
\begin{equation}
\left[\frac{\hat{p}^2}{2} +V(x) \right]  \left| \phi_{n\kappa(t)}
\right>=\varepsilon_{n\kappa(t)}  \left| \phi_{n\kappa(t)}
\right>.\label{eq:Bloch-Houston}
\end{equation}
They can be viewed as adiabatic Bloch states, with $\kappa(t)$
replacing the Bloch-state crystal momentum $k$. ${\kappa(t)}$ thus
parameterizes the field-driven electronic evolution under the
influence of the IR pulse out of an initial state with crystal
momentum $k$ (Fig.~\ref{fig:fig01}). As for ordinary Bloch
functions~\cite{ashcroft}, Houston states with different initial
(field-free) crystal momentum $k$ or different band indices $n$ are
not coupled by the Hamiltonian in Eq.~(\ref{eq:Bloch-Houston}), yet
evolve differently. Since Houston states for different initial
momenta $k$ explore the electronic band $n$ distinctively, they will
be referred to as ``$k-$channels'' in this work.

\subsubsection{\label{subsubsec:densmat}Density-matrix formulation}

The electronic evolution described in Eq.~(\ref{houstonx_c}) can
also be expressed in terms of the density operator,
\begin{equation*}
\hat{\rho}(t)=\left|\psi_k(t)\right>\left<\psi_k(t)\right|\,,
\end{equation*}
in the pure state ${\left| \psi_k (t)\right>}$ given by
Eq.~(\ref{eq:wavefu}). The matrix elements
\begin{equation*}
\begin{aligned}
\rho^{nn'}_k(t)&=B_{nk}(t)B_{n'k}^* (t)   \\
&=\left<\phi_{n\kappa(t)}\right|\hat{\rho}_k\left|\phi_{n'\kappa(t)}\right>
e^{i  \int_{0}^t \Delta\varepsilon_{\kappa(t')}^{nn'}} dt'
\end{aligned}
\end{equation*}
represent band populations for ${n=n'}$. Their time evolution is
obtained by solving
\begin{align}
i\dot{\rho}^{nn'}_k(t) &= E(t)\sum_{n''}[ D^{n''n}_{\kappa(t)} e^{i \int_{0}^t \Delta\varepsilon_{\kappa(t')}^{nn''} dt'} \rho^{n''n'}_k(t) \nonumber \\
&- D^{n'n''}_{\kappa(t)} e^{i \int_{0}^t
\Delta\varepsilon_{\kappa(t')}^{n''n'} dt'}\rho^{nn''}_k(t)].
\label{eq:dmf}
\end{align}

\subsubsection{\label{subsubsec:intrainter}Intra- and interband yield}

The electronic current in each ${k-}$channel,
\begin{equation*}
J_k(t)=J_k^{ra}(t)+J_k^{er}(t) ,
\end{equation*}
consists of intra- and interband contributions,
\begin{align}
J_k^{ra}(t)&=-\sum_n \rho^{nn}_{k}(t)\,P^{nn}_{\kappa(t)}
\label{j_intra}
\end{align}
and
\begin{align}
J_k^{er}(t)&=-\sum_{n'> n}\sum_n  e^{i  \int_{0}^t \Delta\varepsilon_{\kappa(t')}^{nn'} dt'}\,  %\\
\rho^{n'n}_k (t) P^{nn'}_{\kappa(t)} + \text{c.c.}, \label{j_inter}
\end{align}
respectively. It defines the spectral HHG yield from a given
$k-$channel,
\begin{equation}
\begin{aligned}
Y_k(\omega)&=\left|\int_{-\infty}^{\infty} dt \, e^{-i \omega t} J_k(t)\right|^2 \equiv \left|\hat{J}_k(\omega)\right|^2 \\
&=Y_k^{ra}(\omega)+Y_k^{er}(\omega)+2\hat{J}_k^{er}(\omega)\hat{J}_k^{ra}(\omega),
\end{aligned}
\label{eq:yields}
\end{equation}
which, in addition to the intra- [$Y_k^{ra}(\omega)$] and interband
[$Y_k^{er}(\omega)$] yields, includes the interference term
$2\hat{J}_k^{er}(\omega)\hat{J}_k^{ra}(\omega)$.

For the dielectric solid analyzed in this work, the Fermi energy
lies in the band gap between the highest occupied band (the
valence-band) and the lowest unoccupied band (the first conduction
band). Since the  valence-band is fully occupied, we obtain the
total HHG yield as
\begin{equation}
Y\left(\omega\right)=\left| \int_{-\infty}^{\infty} dt \, e^{-i
\omega t} \int_{BZ} dk\, J_{k}(t)\right|^2, \label{eq:yield_deff}
\end{equation}
and corresponding expressions for the total intra-
[${Y^{ra}(\omega)}$] and interband HHG yields [${Y^{er}(\omega)}$],
after including current contributions from all ${k-}$channels,
i.e.~, from all crystal momenta $k$ in the first BZ.

\subsection{\label{subsec:two_band} HHG mechanism for a two-band system}

In this subsection, we restrict the theory developed in
Sec.~\ref{subsec:tdse} to two electronic bands: the valence and
first conduction band, even though more than two bands are required
and included in our converged numerical calculations in
Sec.~\ref{sec:results} below. Designating the valence and conduction
bands with superscripts $v$ and $c$, respectively, the  intra- and
interband  currents in Eqs.~(\ref{j_intra}) and (\ref{j_inter})
simplify to
\begin{eqnarray}
J_k^{ra}(t)&=&-\rho^{vv}_{k}(t)\,P^{vv}_{\kappa(t)} -\rho^{cc}_{k}(t)\,P^{cc}_{\kappa(t)} \label{eq:intra_two}\\
J_k^{er}(t)&=&- e^{-i S(k,\,t) } \rho^{cv}_k(t) P^{vc}_{\kappa(t)} +
\text{c.c.}\,. \label{eq:inter_two_a}
\end{eqnarray}
As detailed in Appendix~\ref{app:aux}, Eq.~(\ref{eq:inter_two_a})
can be written as
\begin{eqnarray}
J_k^{er}(t)&=&\frac{d}{dt}\left[\rho^{cv}_k(t) D^{vc}_{\kappa(t)}e^{-i S(k,\,t)}\right]
 -\Delta J_k^{er}(t) + \text{c.c.} \nonumber \\
 &=&\bar{J}^{er}_k(t)+\text{c.c.}
 , \label{eq:inter_two}
\end{eqnarray}
where we define
\begin{eqnarray}
\Delta J_k^{er}(t) &=& \rho^{cv}_k(t) \dot{D}^{vc}_{\kappa(t)}e^{-i
 S(k,\,t)}\,, \label{eq:DeltaJK} \\
\bar{J}^{er}_k(t) &=& \frac{d}{dt}\left[\rho^{cv}_k(t)
D^{vc}_{\kappa(t)}e^{-i S(k,\,t)}\right] - \Delta J_k^{er}(t)\,,
\end{eqnarray}
and the action
\begin{equation*}
S(k,\,t) = \int_{0}^{t} \Delta\varepsilon_{\kappa(t')}^{cv} dt'.
\end{equation*}
Applying Eq.~(\ref{eq:dmf}) to the two-band system, we obtain
\begin{equation}
i\dot{\rho}^{cv}_k(t)=e^{i  S(k,\,t)} \Delta\rho_k^{vc}(t) E(t)
D^{vc}_{\kappa(t)}, \, \label{eq:polarization}
\end{equation}
with the population difference
\begin{equation}
\Delta\rho_k^{vc}(t)=\left[\rho^{vv}_{k}(t)-\rho^{cc}_{k}(t)\right],
\end{equation}
and
\begin{equation}
i\dot{\rho}^{cc}_k(t)=E(t)D^{vc}_{\kappa(t)} e^{-i
S(k,\,t)}\rho^{cv}(t) -\text{c.c.}\,. \label{eq:population}
\end{equation}

Both, the SBE model for HHG in
Refs.~\cite{vampa2014prl,jiang2017pra,li2018pra} and the expansion
of TSDE solutions in a Houston basis, use an adiabatic basis. Note
that the SBE model in
Refs.~\cite{vampa2014prl,jiang2017pra,li2018pra} adds the damping
term ${\rho^{cv}_k(t)/T_{2}}$ to Eq.~(\ref{eq:polarization}), with
an adjustable damping time $T_{2}$ for the interband
coupling~\cite{haug1998}. In addition, it does not include the term
${\Delta J_k^{er}(t)}$ in Eq.~(\ref{eq:inter_two}). Therefore, the
SBE model~\cite{vampa2014prl} and our SAE-TSDE cannot be expected to
yield identical HHG spectra, not even for ${T_2\rightarrow \infty}$.

\subsubsection{\label{subsubsec:intra_yield}Intraband yield}

For a two-band system, norm preservation demands
${\rho^{vv}_{k}(t)=1-\rho^{cc}_{k}(t)}$, such that the intraband
current in Eq.~(\ref{j_intra}) simplifies to
\begin{equation}
J_k^{ra}(t)=- P^{vv}_{\kappa(t)} + \rho^{cc}_{k}(t)\,\Delta
P^{vc}_{\kappa(t)}, \label{eq:intra_k_curr}
\end{equation}
resulting in the interband yield
\begin{equation}
Y_k^{ra}(\omega)=\left|\int_{-\infty}^{\infty}dt\,e^{-i\omega
t}\left[P^{vv}_{\kappa(t)} - \rho^{cc}_{k}(t)\,\Delta
P^{vc}_{\kappa(t)}\right]\right|^2, \label{eq:intra_k_yield}
\end{equation}
where
\begin{equation}
\Delta P^{vc}_{\kappa(t)} = P^{vv}_{\kappa(t)}-P^{cc}_{\kappa(t)}.
\end{equation}
At low laser intensities, the conduction-band population is very
small (${\rho^{cc}_{k}(t) \ll 1}$), and, for every ${k-}$channel,
the by far dominant contribution to the intraband HHG spectrum is
generated by the term ${P^{vv}_{\kappa(t)}}$ in the intraband
current. With increasing driving-field intensities, the second term
(${\sim \rho^{cc}_k}$) gains significance.

Upon integration over the first BZ, the first term in
Eq.~(\ref{eq:intra_k_curr}) vanishes, due to symmetry, since
${P^{vv}_k}$ is an odd function of $k$ (Fig.~\ref{fig:fig01}). The
integrated current is therefore given by
\begin{equation}
J^{ra}(t)=\int_{BZ} dk\,\rho^{cc}_{k}(t)\Delta P^{vc}_{\kappa(t)}\,,
\label{eq:intra_bz_curr}
\end{equation}
leading to the intraband HHG yield for the two-band system
\begin{equation}
Y^{ra}(\omega)=\left|\int_{-\infty}^{\infty}dt\,e^{-i\omega
t}\int_{BZ} dk\,\rho^{cc}_{k}(t)\Delta
P^{vc}_{\kappa(t)}\right|^2\,. \label{eq:intraband-yield}
\end{equation}
At low intensities, even though each ${k-}$channel defines a
significant intraband current, due to the symmetry-related
cancellation of contributions from crystal momenta ${+k}$ and
${-k}$, the integrated intraband HHG yield is comparatively small.

\subsubsection{\label{subsubsec:inter_yield}Interband yield}

Calculation of the HHG yield for a specific $k$-channel, according to
Eq.~({\ref{eq:yields}}),  requires the
Fourier-transformed current
\begin{equation}
\begin{aligned}
\hat{\bar{J}}^{er}_k(\omega)&=i\omega\int_{-\infty}^\infty dt\,
e^{-i\omega t}\rho^{cv}_k(t) D^{vc}_{\kappa(t)}e^{-i  S(k,\,t)}  \\
&-\int_{-\infty}^\infty dt\, e^{-i\omega t} \Delta J_k^{er}(t) \,.
\end{aligned}
\label{eq:inter3}
\end{equation}
Replacing ${\rho^{cv}_k(t)}$, obtained by integrating
Eq.~({\ref{eq:polarization}}), in Eq.~({\ref{eq:inter_two}}), the
interband current and HHG yield for the two-band system become
\begin{equation}
\begin{aligned}
\hat{\bar{J}}^{er}_k(\omega)&=\omega\int_{-\infty}^\infty dt\,
e^{-i\omega t}D^{vc}_{\kappa(t)}\,e^{-i  S(k,\,t)} \\
&\times \int_0^t dt'\,e^{i  S(k,\,t')} \Delta\rho_k^{vc}(t') E(t') D^{vc}_{\kappa(t')}  \\
&+\int_{-\infty}^\infty dt\, e^{-i\omega t} \Delta J_k^{er}(t)
\end{aligned}
\label{eq:inter4}
\end{equation}
and
\begin{equation}
Y_k^{er}(\omega) = \left|\hat{J}^{er}_k(\omega)\right|^2 =
\left|\hat{\bar{J}}^{er}_k(\omega)+\hat{\bar{J}}^{er}_k(-\omega)^*\right|^2\,,
\end{equation}
respectively.

\subsection{\label{subsec:ApproxInterbandYield}Approximate evaluation of the  interband yield}

While the numerical HHG spectra discussed in Sec.~\ref{sec:results}
below are calculated based on the theory outlined in
Secs.~\ref{subsec:tdse} and \ref{subsec:two_band}, we apply in this
subsection additional approximations to the interband current in
order to derive analytical expressions that reveal additional
physical properties of the interband HHG process in solids. We
restrict this analysis to vector-potential amplitudes ${A_0<\pi/a}$,
for which the range of ${\kappa(t)}$ is limited by the width,
${2\pi/a}$, of one BZ (Fig.~\ref{fig:fig01}).

\subsubsection{\label{subsubsec:stat-phase-approx}Stationary-phase approximation}

Setting the TDM in Eq.~(\ref{eq:inter4}) equal to its value at the
band center (${k=0}$)~\cite{vampa2014prl,vampa2015prb},
${D^{vc}_{\kappa(t)}\approx D^{vc}_0}$, remembering that according
to Eq.~(\ref{eq:tdm})
${\left(D^{vc}_0\right)^2=-\left|D^{vc}_0\right|^2}$, and applying
the frozen valence-band approximation~\cite{keldysh1965jetp}
${|\Delta\rho_k^{vc}(t)|\approx 1}$~\cite{keldysh1965jetp},
Eq.~(\ref{eq:inter4}) simplifies to
\begin{equation}
\begin{aligned}
\hat{\bar{J}}^{er}_k(\omega)\approx &-\omega
\left|D^{vc}_0\right|^2\int_{-\infty}^\infty dt\, \int_0^t dt'\,e^{i
\sigma_\omega(k,\,t,\,t')}  E(t') \,,
\end{aligned}
\label{eq:inter5}
\end{equation}
where
\begin{equation}
%\begin{aligned}
\sigma_\omega(k,\,t,\,t')
%  &=-\omega t-S(k,\,t)+S(k,\,t')  \\
=-\omega t-\int_{t'}^t \Delta\varepsilon_{\kappa(t'')}^{cv} dt'' \,.
%\end{aligned}
\label{eq:full_action}
\end{equation}

Since ${e^{i\sigma_\omega(k,\,t,\,t')}}$ rapidly oscillates as a
function of $t$ and ${t'}$, the integrals in Eq.~(\ref{eq:inter5}) are
dominated by contributions at times ${t=t_e}$ and ${t'=t_s}$ when the
phase ${\sigma_\omega(k,\,t,\,t')}$ is stationary, i.e.,
\begin{equation*}
 \frac{\partial  \sigma_\omega(k,\,t_e,\,t_s)}{\partial t}= \frac{\partial  \sigma_\omega(k,\,t_e,\,t_s)}{\partial
 t'}=0\,.
\end{equation*}
At these times we find
\begin{eqnarray}
 \Delta\varepsilon_{\kappa(t_e)}^{cv} &= \omega \label{S_b}\\
 \Delta\varepsilon_{\kappa(t_s)}^{cv} &= 0 \label{S_a}
 \,.
\end{eqnarray}
Evaluating Eq.~(\ref{eq:inter5}) in stationary-phase
approximation~\cite{lew1994pra,vampa2014prl,wong2001} now results in
\begin{equation}
\hat{\bar{J}}^{er}_k (\omega)\approx -\omega
\left|D^{vc}_0\right|^2\sum_{t_e}\sum_{t_s}\frac{ \left(2\pi
i\right) E(t_s) e^{i \sigma_\omega(k,\,t_e,\,t_s)} }{\left|
\text{det}\left[\text{Hess }
\sigma_\omega(k,\,t_e,\,t_s)\right]\right|^{1/2}} \,,
\label{eq:yield_a}
\end{equation}
with the Hessian matrix
\begin{equation*}
\left[\text{Hess } \sigma_\omega(k,\,t_e,\,t_s)\right]_{ij} = \left(
\frac{\partial^2 \sigma_\omega(k,\,t_e,\,t_s)}{\partial t_i \partial
t_j}\right)_{t_i,t_j=\,t_e,\,t_s} \,.
\end{equation*}
Equation~(\ref{S_b}) implies that for each ${k-}$channel the cutoff
frequency for interband HHG becomes
\begin{equation}
\omega^c(k,\,A_0)=\left \{
  \begin{tabular}{lc}
  $\Delta\varepsilon_{|k|+A_0}^{cv}$ &\;, $|k|+A_0<\pi/a\;\,$ \\
  $\Delta\varepsilon_{\pi/a}^{cv}$&\;, $|k|+A_0\geq \pi/a.  \;$
  \end{tabular}
  \right.
\label{eq:egmax}
\end{equation}

Equation~(\ref{S_a}) cannot be fulfilled in our Kronig-Penney model
for real-valued times and energies because the bandgap is nonzero
across the entire BZ. We designate  the maximal cutoff energy as
${\Delta\varepsilon_{max}^{cv}}$. Complying with Eq.~(\ref{S_a})
requests allowing for complex-valued times, energies, and crystal
momenta $k$. Designating the complex crystal momentum as $K$, we
analytically continue Eq.~(\ref{S_a}) and the transcendental
Eq.~(\ref{eq:kp}) into the complex $K-$plane. Even though complex
roots ${K_s=K(t_s)}$ of Eq.~(\ref{S_a}) can only be obtained
numerically, we get further insight into the interband HHG process
by Taylor-expanding about ${K=0}$,
\begin{equation*}
\Delta\varepsilon_{K}^{cv}
=\Delta\varepsilon_{0}^{cv}+\frac{\text{d}
\Delta\varepsilon_{0}^{cv}}{\text{d} K}
K+\frac{1}{2}\frac{\text{d}\,^2 \Delta\varepsilon_{0}^{cv}}{\text{d}
K^2}K^2 + \mathcal{O}(K^3) \,.
\end{equation*}
While the expansion coefficients are in general complex, we show in
Appendix~\ref{app_an_cont} that ${\frac{\text{d}
\Delta\varepsilon_{0}^{cv}}{\text{d} K}=0}$ and ${\Im
\left[\frac{\text{d}^2 \Delta\varepsilon_{0}^{cv}}{\text{d}
K^2}\right]=0}$, such that
\begin{equation}
\Delta\varepsilon_{K}^{cv}=\Delta\varepsilon_{0}^{cv}+\frac{K^2}{2m^*_0}
+ \mathcal{O}(K^3) \,, \label{eq:egK}
\end{equation}
where $\Im$ stands for ``imaginary part of''. We evaluate the
real-valued reduced effective mass,
\begin{equation*}
m_k^* = \frac{ m_{vk}^*m_{ck}^*}{m_{vk}^* - m_{ck}^*}\,,
\end{equation*}
at the band center,
\begin{equation*}
m_0^* =\left.m_k^*\right|_{k=0} \,,
\end{equation*}
in terms of the valence- and conduction-band effective masses
\begin{equation}
\frac{1}{m_{nk}^*}=\frac{\partial^2 \varepsilon_{nk}}{\partial k^2}
\;,\;\;n=v,\,c \,.  \label{eq:Inversemass}
\end{equation}
Note that the derivatives in Eq.~(\ref{eq:Inversemass}) are taken
along the real axis and ${ \varepsilon_{nk}= \Re
[\varepsilon_{nK}]}$, where $\Re$ stands for ``real part of''. Since
the first term and the coefficient of the quadratic term in
Eq.~(\ref{eq:egK}) are real, the roots $K_s$ are purely imaginary
and correspond to interband transitions at the $\Gamma-$point
(${k=0}$).

In Appendix \ref{app:saddle_a} we show that, for a continuous-wave
driving field of the form ${A_0 \sin(\omega_0 t)}$ and for
${\omega_0 \sqrt{2 \Delta\varepsilon_{0}^{cv} m_0^
*}<E_0}$, the interband HHG yield is given by
\begin{equation}
\begin{aligned}
Y^{er}_k (\omega)&\approx \exp\left[-\frac{2\,\sqrt{2
m^*_0}\,{\Delta\varepsilon_{0}^{cv}}^{3/2}}{E_0\sqrt{1-(k/A_0)^2}}\right]
  \\
&\times\left(2\pi\omega \right)^2\left|D^{vc}_0\right|^4 E_0
\sqrt{1-(k/A_0)^2}
\sqrt{\frac{m^*_0}{2\Delta\varepsilon_{0}^{cv}}}  \\
&\times \left|\sum_{t_e} \frac{e^{-i\omega t_e}\left[ e^{i\left\{\Re
[S(k,\,t_s)]+
S(k,\,t_e)+\frac{\pi}{2}\right\}}+\text{c.c.}\right]}{\left|
E(t_e)\Delta P^{vc}_{\kappa(t_e)}\right|^{1/2}}\right|^2 \,.
\end{aligned}
\label{eq:bc_approx2}
\end{equation}
Applying the effective mass theorem~\cite{ashcroft},
\begin{equation*}
\frac{1}{m^*_{nk}}=1+2\sum_{n'\neq n} \frac{(P_{k}^{nn'})^2}{\Delta
\varepsilon_{k}^{nn'}} \,,
\end{equation*}
we obtain
\begin{equation}
m^*_0 \approx
\frac{1}{4\,\Delta\varepsilon_{0}^{cv}\left|D_{0}^{vc}\right|^2}\,.
\label{eq:pvc_m}
\end{equation}
This allows us to rewrite Eq.~(\ref{eq:bc_approx2}) as
\begin{equation}
\begin{aligned}
Y^{er}_k (\omega)&\approx \exp\left[-\frac{\sqrt{2}\,
\Delta\varepsilon_{0}^{cv}}{E_0 \left|D_{0}^{vc}\right|\sqrt{1-(k/A_0)^2}}\right] \\
&\times2\left(\pi\omega \left|D^{vc}_0\right|\right)^2 \left(\frac{E_0 \left|D_{0}^{vc}\right|
\sqrt{1-(k/A_0)^2}}{\sqrt{2}\Delta\varepsilon_{0}^{cv}}\right)  \\
&\times \left|\sum_{t_e} \frac{e^{-i\omega t_e}\left[ e^{i\left\{\Re
[S(k,\,t_s)]+
S(k,\,t_e)+\frac{\pi}{2}\right\}}+\text{c.c.}\right]}{\left|
E(t_e)\Delta P^{vc}_{\kappa(t_e)}\right|^{1/2}}\right|^2.
\end{aligned}
\label{eq:bc_approx3}
\end{equation}
As expected, the interband HHG yield increases with decreasing band
gap ${\Delta\varepsilon_{0}^{cv}}$ and increasing  TDM
${\left|D^{vc}_0\right|}$.

\subsubsection{\label{subsubsec:relevant-k-range}Relevant k-range for the interband HHG}

The repeated-zone scheme in Fig.~\ref{fig:fig01} shows that, for
${k>A_0}$ or ${k<-A_0}$, $0 < |\kappa(t)| < 2\pi/a$. Hence,
$|\kappa(t)|$ does not reach the limits $0$ and $2\pi/a$,
Eq.~(\ref{S_a}) cannot be satisfied, and contributions to HHG are
restricted to the interval ${k\in (-A_0,\,A_0)}$ within the first
BZ. For the symmetrical dispersion relation of the Kronig-Penney
model (Fig.~\ref{fig:fig01}), the interband HHG yield in
Eq.~(\ref{eq:bc_approx3}) is largest at the $\Gamma-$point.

With the upper limit for the integrated interband yield
\begin{equation}
Y^{er}(\omega)=\left|\int_{BZ} dk\, \hat{J}^{er}_k(\omega)\right|^2
\leq  \left(\int_{BZ} dk\,
\left|\hat{J}^{er}_k(\omega)\right|\right)^2
\end{equation}
and since ${\left[1-(k/A_0)^2\right]^{1/2}\leq 1}$ in
Eq.~(\ref{eq:bc_approx3}), we find the contribution of each
$k$-channel to the interband yield to be limited by
\begin{eqnarray}
\left|\hat{J}^{er}_k(\omega)\right| &=&
\sqrt{Y^{er}_k(\omega)} \nonumber \\
&\propto& \,\exp{\left[-\frac{ \Delta\varepsilon_{0}^{cv}}{\sqrt{2}
E_0 \left|D_{0}^{vc}\right|\sqrt{1-(k/A_0)^2}}\right]}\,.
\end{eqnarray}

We can now estimate the range ${k_{max}(A_0)}$ of initial-state
crystal momenta ${k\in
\left(-\frac{k_{max}(A_0)}{2},\frac{k_{max}(A_0)}{2}\right)}$ that
yield relative contributions larger than ${10^{-N}}$ times the
maximal yield $Y^{er}_0(\omega)$,
\begin{equation*}
Y^{er}_k(\omega) \geq 10^{-N} Y^{er}_0(\omega),
\end{equation*}
as
\begin{equation}
\begin{aligned}
k_{max}(A_0)&\approx\frac{2E_0}{\omega_0} \\
&\times \sqrt{1-\left[\frac{ \Delta\varepsilon_{0}^{cv}}{\sqrt{2}
E_0 \left|D_{0}^{vc}\right|\left( \frac{N}{2}\ln 10+\frac{\sqrt{2}\,
\Delta\varepsilon_{0}^{cv}}{E_0
\left|D_{0}^{vc}\right|}\right)}\right]^2} \,.
\end{aligned}
\label{eq:deltak}
\end{equation}
%and therefore which range ${k_{max}(A_0)}$ we should use in numerical
%calculations when it is not possible to calculate the spectra for
%the entire BZ.

\subsubsection{\label{subsubsec:inter_yield_bz} Integrated interband yield}

In this subsection we examine the net contribution to the interband
current from all $k$-channels in the first BZ to obtain the
observable integrated interband yield ${Y^{er}(\omega)}$. For
numerical applications, the numerical effort can possibly be reduced by
limiting the integration range to ${k_{max}(A_0)}$, as determined in
the previous subsection. Whether this is possible depends on the
specific laser parameters and solid electronic structure.

Equation~(\ref{eq:inter3}) leads to the interband current,
integrated over the whole BZ,
\begin{equation}
\begin{aligned}
\hat{\bar{J}}^{er}(t) &= \int_{BZ}dk \left[ \omega\int_{-\infty}^\infty dt\, e^{-i\omega t}\rho^{cv}_k(t) D^{vc}_{\kappa(t)} e^{-iS(k,\,t)}\right.  \\
&\left.- \int_{-\infty}^\infty dt\, e^{-i\omega t} \Delta
J_k^{er}(t) \right] \,.
\end{aligned}
\end{equation}
For constant TDMs and a frozen valence-band population, we arrive at
the saddle-point conditions
\begin{equation*}
 \frac{\partial  \sigma_\omega(k_r,\,t_s,\,t_e)}{\partial k}=\frac{\partial  \sigma_\omega(k_r,\,t_s,\,t_e)}{\partial t}= \frac{\partial  \sigma_\omega(k_r,\,t_s,\,t_e)}{\partial t'}=0\,,
\end{equation*}
with $ {\sigma_\omega(k,\,t',\,t)}$ defined in
Eq.~(\ref{eq:full_action}). These conditions imply
\begin{eqnarray}
\Delta\varepsilon_{\kappa_r(t_s)}^{cv} &=&0 \label{S_a2}  \\
\int_{t_s}^{t_e}\Delta P^{vc}_{\kappa_r(t)} dt &=& 0 \label{S_c}\\
\Delta\varepsilon_{\kappa_r(t_e)}^{cv} &=& \omega \label{S_b2},
\end{eqnarray}
where  ${\kappa_r(t)=k_r+A(t)}$, and  allow us to determine the
roots $k_r$, $t_s$, and $t_e$ numerically.

Equations~(\ref{S_a2}) and (\ref{S_b2}) are equivalent to
Eqs.~(\ref{S_a}) and (\ref{S_b}). Condition~(\ref{S_c}) arises due
to the integration over the BZ. It expresses the requirement of the
excited photoelectron wave packet, moving with group velocity
${P^{cc}_{k_r+A(t)}}$, and the residual hole wave packet,
propagating with  group velocity $P^{vv}_{k_r+A(t)}$, to recombine
at time $t_e$ after their birth at time $t_s$, while emitting a
photon with energy ${\Delta\varepsilon_{\kappa_r(t_e)}^{cv}}$. We
show in Appendix \ref{app:saddle_kr} that, since ${\Im [t_s]\propto
\omega_0\sqrt{2\Delta \varepsilon^{cv}_0 m_0^*}}$ and the matrix
elements $P^{nn}_k$ are odd functions of $k$, Eq.~(\ref{S_c}) can be
approximated as
\begin{equation}
\int_{t_0}^{t_e}\Delta P^{vc}_{\kappa_r (t)} dt \approx 0\,.
\label{S_c_real}
\end{equation}
Even though within the Kronig-Penney model the band-gap energy
${\Delta\varepsilon_{k}^{cv}}$ grows continuously from the center to
the edge of the first BZ, the $k-$channel at ${k=k_r^c}$ with the
largest frequency,
\begin{equation}
\omega^{c}_{BZ}(A_0)=\varepsilon_{k_r^c+A(t_e)} \,,
\end{equation}
lies in the range ${k_r\in (-A_0,\,A_0)}$, as seen in
Sec.~\ref{subsubsec:relevant-k-range}, and needs to be determined
numerically (see Sec.~\ref{subsubsec:multi_field_dep} for a specific
numerical example).

In analogy to Eq.~(\ref{eq:yield_a}), we obtain the
Fourier-transformed net interband current as
\begin{equation}
\begin{aligned}
\hat{\bar{J}}^{er} (\omega)&\approx  -\omega \left|D^{vc}_0\right|^2 \, \left(2\pi i\right)^{3/2} \\
& \times \sum_{|k_r|<A_0}\sum_{t_e}\sum_{t_s}\frac{  E(t_s)   e^{i
\sigma_\omega(k_r,\,t_e,\,t_s)} }{\left| \text{det}\left[\text{Hess
} \sigma_\omega(k_r,\,t_e,\,t_s)\right]\right|^{1/2}} \,.
\end{aligned}
\label{eq:yield_d}
\end{equation}
This expression is equivalent to
\begin{equation}
\begin{aligned}
\hat{\bar{J}}^{er} (\omega)&\approx  -\omega \left|D^{vc}_0\right|^2 \, \left(2\pi i\right)^{3/2} \\
&\times \sum_{|k_r|<A_0}\sum_{t_e}\sum_{t_s}  E(t_s)   e^{i \sigma_\omega(k_r,\,t_e,\,t_s)}  \\
&\times y^{cv}_{\kappa_r}(t_e,t_s)\,,
\end{aligned}
\end{equation}
where
\begin{equation}
\begin{aligned}
y^{cv}_{\kappa_r}(t_e,t_s) &=\left| \sqrt{\frac{ 2
\Delta\varepsilon_{0}^{cv}}{m_0^*}}E(t_e)
\Delta P^{vc}_{\kappa_r(t_e)}\right|^{-1/2}\\
&\times \left|\left[\sqrt{\frac{ 2
\Delta\varepsilon_{0}^{cv}}{m_0^*}}
+iE(t_s)\int_{t_s}^{t_e}dt\frac{1}{m^*_{\kappa_r(t)}}\right]\right|^{-1/2}\,,
\end{aligned}
\end{equation}
resulting in the approximated integrated interband yield
\begin{equation}
\begin{aligned}
Y^{er} (\omega)&\approx  (2\pi)^3\left(\omega \left|D^{vc}_0\right|\right)^2  \\
&\times | \sum_{|k_r|<A_0}  E(t_s) \,e^{\left[-\frac{\sqrt{2}\, \Delta\varepsilon_{0}^{cv}}{E_0 \left|D_{0}^{vc}\right|\sqrt{1-(k_r/A_0)^2}}\right]} \, \\
&\times \sum_{t_e} \left[  e^{-i\omega t_e} e^{i\left\{\Re [S(k,\,t_s)]+ S(k,\,t_e)+\frac{3\pi}{2}\right\}}+\text{c.c.}\right]   \\
&\times y^{cv}_{\kappa_r}(t_e,t_s) |^2 \,.
\end{aligned}
\label{eq:bc_approx4}
\end{equation}

\section{\label{sec:results}Numerical Results}

\begin{figure}[h]
\centering{}\includegraphics[width=0.8\linewidth]{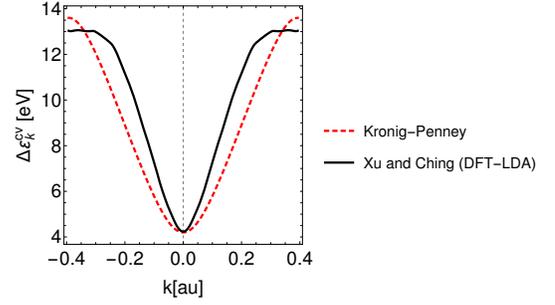}
\caption{(Color online). Bandgap energies in the first BZ between
the valence and lowest conduction band of MgO obtained within the
Kronig-Penny model with interlayer spacing of 8~a.u. and potential
strength  22.345~eV (black solid line) and adapted from the
OLCAO-LDA DFT calculation of Ref.~\cite{xuprb1991} along the
$\Gamma-X$ direction (dashed red line). \label{fig:fig02}  }
\end{figure}

For our numerical applications of the theoretical model described in
Sec.~\ref{sec:theo}, we adopt the laser wavelength (3250~nm) and
pulse duration (10 optical cycles) of the experiment by
Ghimire~\etal~\cite{ghimire2011natphys} and the temporal pulse
profile given by Eq.~(\ref{eq:pulseshape}). We model the electronic
structure of MgO based on the Kronig-Penney model potential
[Eq.~(\ref{eq:KP-pot})] with an interlayer spacing of
${a=8}$~a.u.~\cite{xuprb1991} and adjust the potential strength to
${V_0=22.345}$~eV in order to reproduce the bandgap energy between
the valence and conduction band at the $\Gamma-$point, 4.19~eV,
obtained by Xu and Ching~\cite{xuprb1991} from a full-dimensionality
orthogonalized-linear-combination-of-atomic-orbitals (OLCAO)
calculation within the framework of density-functional theory (DFT)
in local density approximation (LDA). These values of the
Kronig-Penney potential parameters result in a local bandgap at the
BZ edge (at ${k=\pm \pi/a}$) of 13.6~eV, in very good agreement with
the local bandgap at the $X$ point of 13~eV computed by Xu and Ching
(Fig.~\ref{fig:fig02}).

In this section, we perform a systematic numerical study of the
contributions to the HHG spectrum from different $k$ channels within
the entire first BZ. We first discuss results obtained by
restricting the external-field-driven electron dynamics to the
valence and conduction band in Sec.~\ref{subsec:res_two}, before
presenting converged HHG spectra obtained by including up to  13
electronic bands of MgO in Sec.~\ref{subsec:multiband}. For all
calculations we employed a fourth-order Runge-Kutta algorithm to
numerically solve Eq.~(\ref{houstonx_c}) at 400 equally spaced
${k-}$points in the first BZ.

\subsection{\label{subsec:res_two} HHG spectra in two-band approximation}

In order to understand the basic mechanisms of intra- and interband
HHG in solids, we complement our theoretical analysis of HHG by a
two-band system in Sec.~\ref{subsec:two_band} above, with the
numerical solution of Eq.~(\ref{houstonx_c}), restricted to the
lowest (highest) conduction (valence) band of MgO. While this
approach can only provide acceptable results at moderate intensities
of the driving field (${A_0<\pi/a}$), it fails at higher field
strengths, for which the inclusion of more bands is mandatory
(Sec.~\ref{subsec:multiband}).

\subsubsection{\label{subsubsec:k_resolved}Lattice-momentum-resolved contributions to HHG}

\begin{figure}[h!]
\includegraphics[width=1\linewidth]{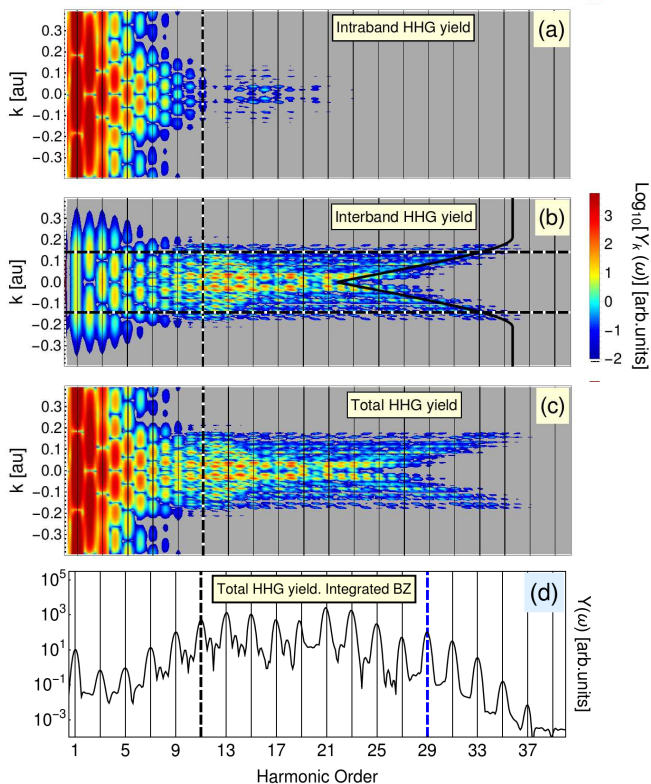}
\caption{(Color online). HHG in MgO driven by a 10-cycle 3250 nm
pulse with a peak electric-field strength of ${0.13\,\text{V/\AA}}$
(corresponding to a peak intensity of $2.24\times 10^{11}$
W/cm$^2$).
Thin vertical lines mark odd harmonics. The vertical black dashed
line indicates the harmonic order ($\Delta\varepsilon_{0}^{cv}
/\omega_0 = 11$) corresponding to the minimum band-gap energy
$\Delta\varepsilon_0^{cv}$ at $k=0$.
(a) Intraband HHG spectrum.
(b) Interband HHG spectrum. The horizontal dashed black lines
indicate ${\pm k_{max}(A_0)/2}$ for $N=5$ as given by
Eq.~(\ref{eq:deltak}). The V-shaped black line shows the maximal
band-gap energy $\omega^c(k,\,A_0)$ as given by
Eq.~(\ref{eq:egmax}).
(c) Total HHG spectrum.
The yields in (a-c) are given as functions of the lattice momentum
$k$ over the entire first BZ and on the same logarithmic scale.
(d) Total HHG yield, integrated over the first BZ. Above harmonic
order 29, the yield drops exponentially, determining the cutoff
energy indicated by the vertical blue dashed line.
} \label{fig:fig03}
\end{figure}

To reveal the characteristics of HHG spectra, including all
$k$-channels in the first BZ, we performed a calculation at a field
strength of ${0.13\,\text{V/\AA}}$. Figures~\ref{fig:fig03} (a) -
\ref{fig:fig03} (c) display two-band HHG spectra as a function of
the lattice momentum $k$. While, according to Eq.~(\ref{eq:yields}),
the total yield is not equal to the sum of the intra- and interband
yield, it is instructive to examine intra- and interband spectra
separately. The comparison of Figs.~\ref{fig:fig03}(a) and
\ref{fig:fig03}(b) shows that intraband harmonics dominate HHG at
harmonic photon energies below the band gap,
${\Delta\varepsilon_{0}^{cv} \approx 11 \omega_0}$, while above this
threshold mainly interband harmonics contribute to the total HHG
yield in Fig.~\ref{fig:fig03}(c). The observation that the band gap
establishes a threshold between intra- and interband HHG is
consistent with energy conservation, requiring electronic
probability density to cross the local band gap,
${\Delta\varepsilon_{\kappa(t)}^{cv}}$, before contributing to the
interband current and thus to the interband HHG.

\begin{figure}[h]
\includegraphics[width=.6\linewidth]{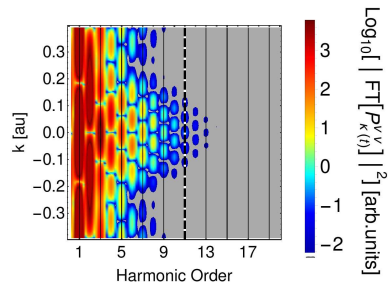}
\caption{(Color online). Time Fourier transformation of the
valence-band group velocity ${P^{vv}_{\kappa(t)}}$ for a 10-cycle
3250 nm laser pulse of ${0.13\, \text{V/\AA}}$ peak field strength.
} \label{fig:fig04}
\end{figure}

In view of Eq.~(\ref{eq:intra_k_curr}) and the semiclassical
interpretation of $P^{vv}_{\kappa(t)}$ as the photoelectron group
velocity in the valence band, at low electric-field strengths we
expect the intraband yield [Eq.~(\ref{eq:intra_k_yield})] to be
dominated by the Fourier transform of $P^{vv}_{\kappa(t)}$
(Fig.~\ref{fig:fig04}). Indeed, below the interband-gap threshold
(11th harmonic), the yields in Figs.~\ref{fig:fig03}(a) and
\ref{fig:fig04} very closely resemble each other and are identical
with respect to their zeros along the $k$-axis for every given
harmonic order: both spectra exhibit even and odd harmonics that
vary over the first BZ. At the $\Gamma-$point and edge of the first
BZ both spectra only include odd harmonics.

Below the interband-gap threshold, the interband spectrum in
Fig.~\ref{fig:fig03}(b) bears some similarity with the intraband
yield in Fig.~\ref{fig:fig03}(a), but has significantly lower yields
and a different distribution of yield nodes along the $k$-axis, at
all harmonic orders. Above the interband-gap threshold, the
interband yield has a rich $k$-dependent structure of even and odd
harmonics. According to Eq.~(\ref{eq:egmax}), the spectral range of
interband high-order harmonics above the interband-gap threshold at
the 11th harmonic is limited by ${\omega^c(k,\,A_0)}$ . This
$k$-dependent upper limit is indicated in Fig.~\ref{fig:fig03}(b) by
the V-shaped solid black line. According to Eq.~(\ref{eq:deltak}),
contributions to the interband yield from $k$-channels that are $N$
orders of magnitude smaller than the maximal yield at the
$Gamma$-point lie in the range ${\pm k_{max}(A_0)/2}$. The V-shaped
solid black line shows their onset for ${N=5}$.

We note that our numerical yields in Fig.~\ref{fig:fig03}, including
contributions to HHG for $k$-channels within the entire first BZ,
are incompatible with the assumption in previous
studies~\cite{ghimire2012pra,wu2015pra} that only a small part of
the first BZ near the ${\Gamma-}$point contributes to HHG in solids.
Even though for the  one-dimensional model solid investigated here
computing time is not an issue, the limit imposed by
Eq.~(\ref{eq:deltak}), and its numerical validation in
Fig.~\ref{fig:fig03}, in addition to providing physical insight into
the HHG process, is relevant for reducing the computational effort
in multi-band HHG calculations based on a three-dimensional
representation of the solid.

While the $k$-channel-resolved HHG spectrum in
Fig.~\ref{fig:fig03}(c) includes even and odd high-order harmonics
and depends in a rather complex way on the harmonic photon energy
and lattice momentum $k$, including contributions to HHG from the
entire first BZ according to Eq.~(\ref{eq:yield_deff}), results in
the comparatively simple total HHG spectrum shown in
Fig.~\ref{fig:fig03}(d). As expected due to the inversion symmetry
about the $\Gamma-$point of the Kronig-Penney band structure
(Fig.~\ref{fig:fig01}), the total HHG spectrum in
Fig.~\ref{fig:fig03}(d) is strongly dominated by odd harmonics.

\subsubsection{\label{subsubsec:field_dep}Field-strength dependence of HHG spectra}

%In order to continue our analysis of the two-band system, we study
%two limiting cases, namely (i) a calculation considering the
%$\Gamma-$point only, and  (ii) one integrating over the whole BZ.

\begin{figure}[h]
\centering{}\includegraphics[width=1\linewidth]{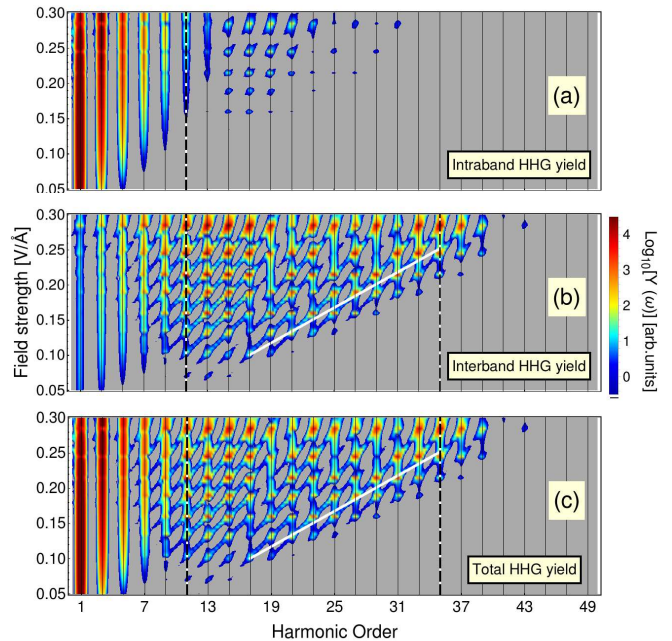}
\caption{(Color online). HHG in MgO driven by a 10-cycle 3250 nm
pulse as a function of peak laser-electric-field strength.
Contributions to the HHG yield from the ${k=0}$ channel ($\Gamma -$point) only.
(a) Intraband, (b) interband, and (c) total spectra.
Thin vertical lines mark odd harmonics. The vertical black dashed
lines correspond to the minimal and maximal local band-gap energies
${\Delta\varepsilon_{0}^{cv}}$ and ${\Delta\varepsilon_{max}^{cv}}$
in the first BZ. The white line in (b,c) indicates the cutoff
harmonic order. } \label{fig:fig05}
\end{figure}

Figure~\ref{fig:fig05} shows the total HHG yield (including intra-
and interband HHG) for the ${k=0}$ channel, i.e., only including the
${\Gamma-}$point at the center of the first BZ, for peak
electric-field strengths
${0.05\,\text{V/\AA}<E_0<0.3\,\text{V/\AA}}$. This field-strength
range corresponds to the laser-peak-intensity range ${3.3\times
10^{10}\,\text{W/cm}^2<\,I_0\,<1.19 \times 10^{12}\,\text{W/cm}^2}$.
We chose the upper limit of this interval to slightly exceed the
field strength ${(\pi/a)\omega_0^{-1}\approx 0.283\,\text{V/\AA}}$.
At this field strength, the vector-potential amplitude is
${A_0=\pi/a}$, such that the field-dressed time-dependent crystal
momentum  ${\kappa(t)=k + A(t)=A(t)}$ defined in
Eq.~(\ref{eq:kappa_deff}) explores the entire first BZ
(Fig.~\ref{fig:fig01}) within one optical cycle in the plateau of
the laser pulse and, thus, the entire range of local band gaps. We
selected the lower limit of the field strength in view of the
approximations made in Sec.~\ref{subsec:two_band}, which are based
on a series expansion in the parameter
${\gamma=\omega_0\sqrt{2\Delta\varepsilon_{0}^{cv}m^*_0}/E_0}$.
Requesting ${\gamma<1}$, implies for MgO ${E_0>0.1\,\text{V/\AA}}$,
slightly above the lower limit of the assumed range of field
strengths.

The comparison of the intra- and interband yields in
Figs.~\ref{fig:fig05}(a) and ~\ref{fig:fig05}(b) shows that the
intraband emission dominates the total yield in
Fig.~\ref{fig:fig05}(c) below and interband emission above the
band-gap threshold near the 11th harmonic, for the entire considered
range of electric-field strengths. The HHG cutoff is thus determined
by interband emission and displayed as superimposed white lines in
Figs.~\ref{fig:fig05}(b) and ~\ref{fig:fig05}(c). We analyze the HHG
cutoff behavior  in more detail in Sec.~\ref{subsubsec:cutoff}
below. Over the range of displayed electric-field strengths the
intra- and interband spectra are dominated by odd harmonics, with
slim traces of even harmonics. Small even and non-integer HHG yields
were also noticed in an SBE-based calculation by
Li~\etal~\cite{li2018pra} and explained in terms of the combined
effect of the external-field and time dependence of the TDM
$D^{vc}_{\kappa(t)}$. We speculate that this effect may be enhanced
due to our inclusion of the term defined by Eq.~(\ref{eq:DeltaJK})
in Eq.~(\ref{eq:inter_two}). This term is absent in the SBE model.

\begin{figure}[h]
\centering{}\includegraphics[width=1\linewidth]{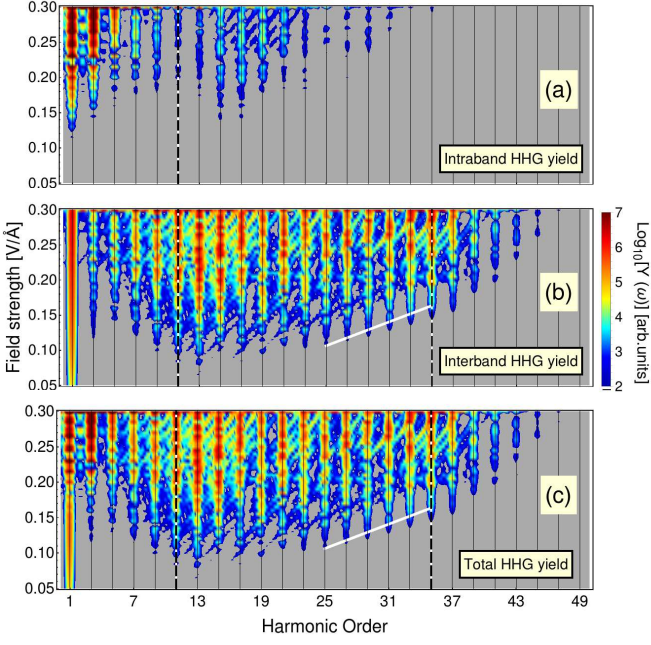}
\caption{(Color online). As Fig.~\ref{fig:fig05}, but integrated
over initial $k$-channels from the entire first BZ.}
\label{fig:fig06}
\end{figure}

Relaxing the ${k=0}$-channel ($\Gamma$-point) emission restriction,
Fig.~\ref{fig:fig06} shows yields obtained after integrating $k$
over the first BZ. As for the $\Gamma$-point-only yield in
Fig.~\ref{fig:fig05}, the comparison of the intra-, interband, and
total yields in Figs.~\ref{fig:fig06}(a), \ref{fig:fig06}(b), and
\ref{fig:fig06}(c), respectively, reveals for the entire shown
laser-peak-intensity range that below the lowest band-gap threshold,
near the 11th harmonic, the yields are dominated by intraband
emission, while above this threshold practically only interband
emission occurs. Even though the spectra include traces of even and
non-integer harmonics (as they do for $\Gamma-$point-only emission),
the contrast between odd and even harmonic yields is larger than for
$\Gamma$-point-only emission.

As discussed in the preceding Sec.~\ref{subsubsec:k_resolved},
intraband emission at the $\Gamma-$point is directly related to the
valence-band group velocity. It yields intense odd and even
harmonics, as seen in Figs.~\ref{fig:fig03} and \ref{fig:fig04}. On
the other hand, as given by Eq.~(\ref{eq:intra_bz_curr}), upon
integration over the first BZ the valence-band term in
Eq.~(\ref{eq:intra_k_yield}) cancels (by symmetry), and high-yield
emission requires the laser-electric field to be strong enough to
effectively promote electrons to the conduction band. This explains
the low yield at low intensities in Figs.~\ref{fig:fig06}(b) and
\ref{fig:fig06}(c), as compared to the corresponding
$\Gamma-$point-only yields in Fig.~\ref{fig:fig05}. The experimental
investigation of HHG below the lowest band-gap threshold and for low
to moderate peak intensities might thus resolve the range
${k_{max}(A_0)}$ of carriers involved in HHG for a specific
substrate. However, it remains to be explored to what extent
experimental focal-volume effects, i.e., averaging of the laser
intensity profile, prevent the accurate determination of
${k_{max}(A_0)}$.

\subsubsection{\label{subsubsec:cutoff}Field-strength dependence of the HHG cutoff}

\begin{figure}[!h]
\centering{}\includegraphics[width=1.0\linewidth]{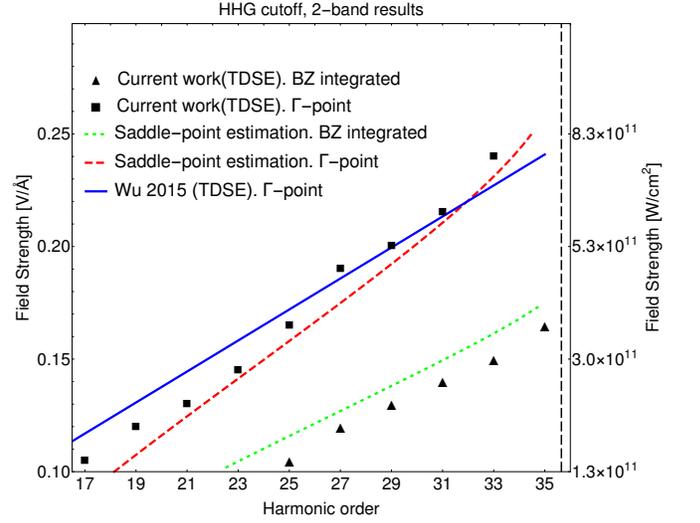}
\caption{(Color online).  Dependence of HHG-cutoff order on the
laser peak-electric-field strength (left vertical axis) or intensity
(right vertical axis) in two-band approximation.
Square markers show  HHG cutoffs for $\Gamma-$point-only (${k=0}$)
emission. Triangular markers indicate HHG cutoffs including
$k$-channels from the entire first BZ.
The red dashed and green solid line show approximate cutoffs
obtained from the saddle-point equations in
Secs.~\ref{subsubsec:stat-phase-approx} and
\ref{subsubsec:inter_yield_bz} for $\Gamma-$point-only emission and
$k$-channels from the entire first BZ, respectively. The dashed red
line, in particular, is given by Eq.~(\ref{eq:egmax}).
The solid blue line shows 2-band TDSE results adapted from Fig.~3(b)
in Wu~\etal~\cite{wu2015pra} for $\Gamma-$point-only emission. }
\label{fig:fig07}
\end{figure}

In this subsection we analyze the field-strength dependence of the
HHG cutoff frequency obtained including both, $\Gamma-$point-only
emission and $k$-channels from the entire first BZ. As the
comparison of interband and total yields in Figs.~\ref{fig:fig05}
and \ref{fig:fig06} reveals, inclusion of the entire first BZ
increases the highest generated frequencies, as compared to
$\Gamma-$point-only emission. For each given peak electric-field
strength, we visually determine the cutoff as the HHG order at which
the yield starts to decline exponentially (as indicated in
Fig.~\ref{fig:fig03}(d)), for both $\Gamma-$point-only and
BZ-integrated yields. This leads to the intensity-dependent cutoff
shown by the markers in Fig.~\ref{fig:fig07}. For the shown range of
harmonic orders, inclusion of all $k$-channels results in cutoffs
(indicated as triangular markers) in good agreement with our
saddle-point prediction (green dotted line) that lie  8 - 10
harmonic orders above the cutoff for $\Gamma-$point-only emission
(square markers). The cutoff orders predicted for the
$\Gamma$-point-only emission by our 2-band TDSE calculations and our
saddle-point analysis (red dashed line) are in fair agreement with
the 2-band TDSE results of Wu~\etal~\cite{wu2015pra} (solid blue
blue line), which we adapted from Fig.~3(b) in
Ref.~\cite{wu2015pra}. In agreement with Wu~\etal, we find that the
cutoff increases approximately linearly with laser
peak-electric-field strength over the displayed range of harmonic
orders, albeit with a noticeably smaller slope.

%  ------------ stationary phase explanation

\begin{figure}[!h]
\centering{}\includegraphics[width=1.0\linewidth]{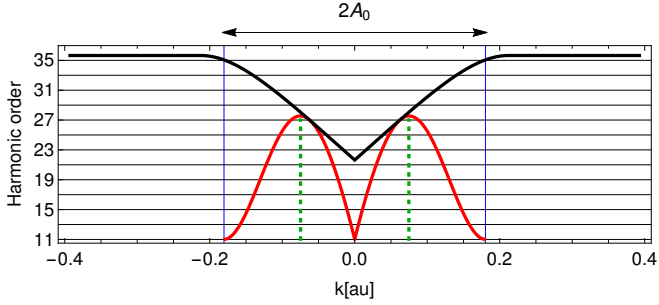}
\caption{(Color online). Determination of the cutoff HHG order for a
driving-laser peak intensity of ${0.13\text{V/\AA}}$ and a model MgO
crystal with interlayer spacing $a=8$~a.u..
The black solid line shows the maximum local vertical band-gap
energy at the field-dressed crystal momentum ${\kappa(t)=k + A(t)}$
during one optical cycle of the driving laser as a function of the
field-free crystal momentum ${k \in [-\pi/a=-0.39,\pi/a=
0.39]}$~a.u..
The vertical thin blue lines indicate crystal momenta ${\pm k_r}$,
determined in stationary-phase approximation, that maximize the
interband HHG yield ${Y^{er}(\omega)}$ given in
Eq.~(\ref{eq:bc_approx4}).
The red line shows the interband HHG cutoff in each $k$ channel.
The vertical green dotted lines indicate the highest cutoff energies
reached at crystal momenta $k_r^c$ that do not need to coincide with
the maximum vertical band gap at the first BZ edge (${k=\pi/a}$).
} \label{fig:fig08}
\end{figure}

To better understand the field-strength dependence of the HHG
cutoff, we resort to the saddle-point analysis of the HHG process in
the 2-band approximation discussed in
Secs.~\ref{subsubsec:stat-phase-approx} and
\ref{subsubsec:inter_yield_bz}. Referring to HHG including all
$k$-channels in the first BZ, we numerically solve the saddle-point
Eq.~(\ref{S_c_real}) for $t_e$ and subsequently Eq.~(\ref{S_b2}) for
each $k_r$. This yields the frequency curve
$\Delta\varepsilon_{\kappa_r(t_e)}^{cv}$, from which we get the
maximal frequency $\omega^c_{BZ}$ and cutoff harmonic order
%$n^c_{BZ} = \omega^c_{BZ}/\omega_0$.
$\omega^c_{BZ}/\omega_0$. The red line in Fig.~\ref{fig:fig08},
shows the harmonic energy ${\Delta\varepsilon_{\kappa_r(t_e)}^{cv}}$
as a function of values for $k_r$ that contribute to the
BZ-integrated interband yield for a peak field strength of the
driving laser of ${0.13\text{V/\AA}}$. As discussed in
Sec.~\ref{subsubsec:inter_yield_bz}, ${|k_r|<A_0}$ (indicated by
vertical thin blue lines). The maximum harmonic energy is obtained
at  crystal momenta ${\pm k_r^c}$ indicated by the dotted green
lines. $k$-channels with ${k=\pm k_r^c}$ thus yield the maximum
cutoff energy at the given laser field strength. The V-shaped black
solid line in Fig.~\ref{fig:fig08} shows the maximum energy
${\omega^c(k,A_0)}$, i.e., the electron-hole-pair-recombination
energy in each $k-$channel given by Eq.~(\ref{eq:egmax}) [cf.,
Fig.~\ref{fig:fig03}(b)]. This energy depends on the  maximum local
band gap in each channel and exceeds the cutoff energy in the
BZ-integrated yield $\Delta\varepsilon_{\kappa_r(t_e)}^{cv}$ (red
line). Since
$\omega^c(k_r,A_0)\approx\Delta\varepsilon_{\kappa_r(t_e)}^{cv}$ the
${k=\pm k_r^c}$ channels have the largest contribution the HHG
yield.

\subsection{\label{subsec:multiband}Multiband spectra}

\subsubsection{\label{subsubsec:multi_field_dep}Field-strength dependence of HHG spectra}

\begin{figure}[h]
\centering{}\includegraphics[width=1\linewidth]{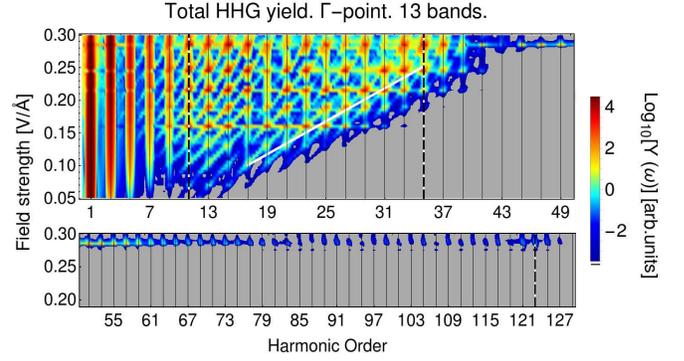}
\caption{(Color online). Total HHG spectrum from MgO driven by a
10-cycle 3250 nm laser. The spectral yield includes intra- and
interband contributions from the ${k=0}$ channel ($\Gamma-$point)
only. Thin vertical lines mark odd harmonics.
(a) Harmonic orders 50 and below. The white line indicates the
cutoff energy.
(b) Harmonic orders 51 to 129.
The vertical black dashed lines correspond to the minimal local
band-gap energy, $\Delta\varepsilon_0^{cv} \approx 11\,\omega_0$,
and maximal local band-gap energies, $\Delta\varepsilon_{max}^{cv}
\approx 35\,\omega_0$ and $\approx 123\,\omega_0$, between the
valence band and lowest and third conduction band, respectively. }
\label{fig:fig09}
\end{figure}

\begin{figure}[h]
\includegraphics[width=1\linewidth]{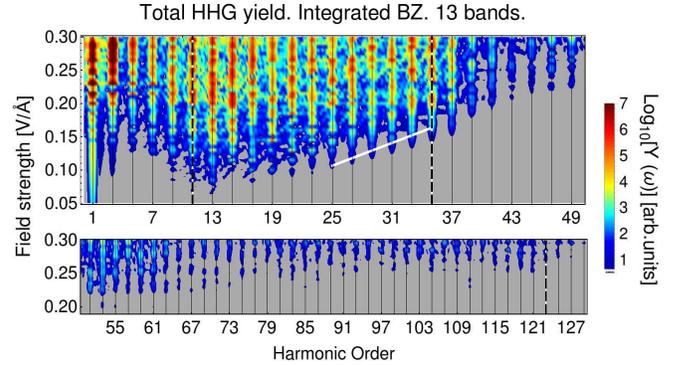}
\caption{(Color online). As Fig.~\ref{fig:fig09}, but integrated
over initial $k$-channels from the entire first BZ.}
\label{fig:fig10}
\end{figure}

Figures~\ref{fig:fig09} and \ref{fig:fig10} show multiband HHG
spectra for both the $\Gamma$-point-only and BZ-integrated
calculations, respectively. Including the lowest 13 bands of the
Kronig-Penney model MgO crystal, we find these spectra to have
converged. The convergence of the results was determined from the
evaluation, in the $k=0$ channel, of the population $\rho^{MM}_0(t)$
in each new band $M$ we added to the calculation. Based on our
convergence criterium $\rho^{MM}_0(t)<10^{-12} \rho^{vv}_0(0)$, we
find $M=13$ for a maximum peak field strength of
${E_0=0.3\,\text{V/\AA}}$. We included the same number (13) of bands
for calculations at lower field strengths and for $k\neq0$.
Comparison with the two-band yields of Figs.~\ref{fig:fig05} and
\ref{fig:fig06} shows that the spectral region below the first
band-gap threshold is dominated by the electron dynamics in the
lowest two bands. For harmonic orders below $\approx 35$, the shape
of the multiband spectra and their laser-electric-field dependence
largely resemble the two-band spectra. Our numerical tests showed
that the inclusion of more than two bands gradually improves the
agreement with fully converged spectra in the shown spectral range.

Above the $\approx$ 35th harmonic and at the highest field strengths
shown in Figs.~\ref{fig:fig09} and \ref{fig:fig10}, a new plateau
emerges, which we attribute to contributions from the second and
third conduction band. If only the ${k=0}$ channel is included, the
second plateau emerges at a higher field strength, $E_0 > \pi
\omega_0/a $, than in calculations including the entire first BZ.
This field strength corresponds to the vector potential for which
the cutoff frequency acquires its maximum value
${\omega^c(0,\,E_0/\omega_0)=\Delta \varepsilon^{vc}_{max}}$ given
by Eq.~(\ref{eq:egmax}).

\subsubsection{\label{subsubsec:cutoff-multi}Field-strength dependence of the HHG cutoff}

\begin{figure}[!h]
\centering{}\includegraphics[width=1.0\linewidth]{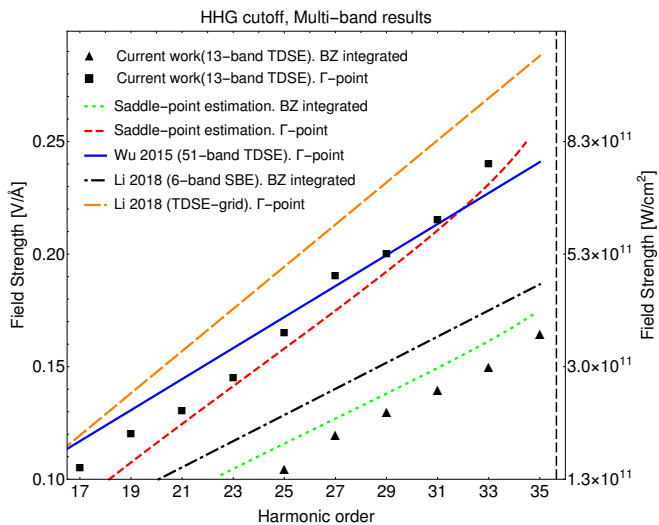}
\caption{(Color online).  Dependence of HHG-cutoff order on the peak
laser-electric-field strength (left vertical axis) or intensity
(right vertical axis) from calculation including more than two
bands.
Square markers show HHG cutoffs for $\Gamma-$point-only (${k=0}$)
emission. Triangular markers indicate HHG cutoffs including
$k$-channels from the entire first BZ.
The blue solid line shows  51-band TDSE results for emission from
the $\Gamma-$point-only, adapted from Fig. 3(a) in
Ref.~\cite{wu2015pra}.
The black dotted and orange dashes lines are adapted from the
BZ-integrated 4-band SBE calculation and the numerical TDSE solution
for $\Gamma-$point-only emission by Li~\etal.
To facilitate the comparison with the 2-band cutoff calculations, we
reproduce the saddle-point-approximation results of
Fig.~\ref{fig:fig07} (red dashed and green solid line).
}
\label{fig:fig11}
\end{figure}

Figure~\ref{fig:fig11} shows the field-strength dependence of the
HHG cutoff we obtained from 13-band TDSE calculations by either
including $\Gamma-$point-only emission (square markers) or
$k$-channels from the entire first BZ (triangular markers), in
comparison with the multi-band calculations from the literature. The
solid blue line shows 51-band TDSE results for emission from the
$\Gamma-$point-only, adapted from Fig. 3(a) in
Ref.~\cite{wu2015pra}. The orange dashed line shows HHG cutoff
orders obtained by the direct propagation of the one-dimensional
TDSE for $\Gamma-$point-only emission by Li~\etal~\cite{li2018pra}.
The black dash-dotted line is adapted from the 6-band SBE
calculation, including $k$-channels from the entire first BZ, of
Li~\etal.

For $\Gamma-$point only emission yields our stationary-phase
approximation results in Fig.~\ref{fig:fig11} (red dashed line)
compare well with our 13-band calculations (square markers). Both
are in reasonable agreement with the 51-band TDSE calculation of
Wu~\etal~\cite{wu2015pra} (blue line). The cutoff harmonic orders we
obtain in stationary-phase approximation are red-shifted by about
three harmonic orders relative to the results of
Li~\etal~\cite{li2018pra} (orange dashed line), but approximately
match the slope of the field-strength-dependent cutoff increase
found by Li~\etal.
The cutoff harmonic orders predicted by our BZ-integrated 13-band
calculation (triangular markers) agree well with our analytical
stationary-phase approximation (green dotted line) and the 6-band
SBE calculation of Li~\etal (black dashed-dotted line). We note that
the calculation of Li~\etal includes a heuristic dephasing time, in
contrast to our approach, where, apart from the adjusted potential
strenghth of the Kronig-Penney model potential, no {\it ad hoc}
parameters are introduced.

\section{Conclusions}

We investigated intra- and interband HHG from a solid within the SAE
approximation, based on a one-dimensional Kronig-Penney model of the
solid's electronic band structure. We expanded the active electron
wavefunction in a basis of adiabatically field-dressed Bloch states
(Houston states). We theoretically analyzed within a
stationary-phase approximation and numerically evaluated in two-band
calculations, contribution to the HHG yield from specific
initial-state crystal momenta $k$ in the first BZ. This revealed
essential contributions to the HHG yield from non-zero crystal
momenta in the first BZ. We found complementary contributions to the
HHG yield from intra- and interband emission. While intraband
emission dominates below the threshold for interband excitation,
interband emission almost exclusively determines the HHG yield above
this threshold. Applying a stationary-phase approximation, we
derived closed analytical expressions for the HHG yields and cutoff
dependence on the laser-electric-field strength. This allow us, for
example, to estimate the loss of accuracy in calculating HHG yields
that are induced by restricting $k$ to a small interval near the
$\Gamma-$point in the first BZ, as compared to BZ-integrated yields.

By adjusting the Kronig-Penney model potential to the published
electronic-structure data for MgO crystals, we validated our
theoretical analysis against numerically calculated $k$-resolved and
$k$-integrated HHG yields and cutoff frequencies for a range of
laser intensities. These analytical predictions and  numerical
calculations are in very good agreement with each other. They are
also in qualitative and fair quantitative agreement with different
numerical calculations from other authors. Investigating the
characteristics of HHG for initial crystal momenta in the entire
first BZ, we find significant HHG yields at even high-order harmonic
orders for ${k\neq0}$ (off the $\Gamma-$point). Even harmonics are
particularly prominent for the lowest intraband harmonics. As
observed experimentally and expected due to the symmetric dispersion
of the model MgO solid with respect to the BZ-zone center, our
BZ-integrated yields predominantly contain odd harmonics. In
contrast to HHG in gaseous atoms, we confirm the cutoff harmonic
order to increase linearly with the peak electric-field strength of
the driving laser.

Our numerical calculations for MgO confirm our
stationary-phase-approximation results that restraining the crystal
momentum to an arbitrarily small range of crystal momenta close to
the $\Gamma-$point does not adequately model HHG in solids driven by
intense laser fields. Strongly depending on the laser intensity, in
calculations with more than two bands, we find a second plateau
above the cutoff HHG order we determined in two-band calculations.

\appendix

\section{\label{app_KP}Kronig-Penney basis functions}

The band structure of the Kronig-Penney model for periodic delta
potentials,
\begin{equation}
V(x) = V_0 \sum_j \delta(x - ja) \,, \label{eq:KP-pot}
\end{equation}
is given by Eq.~(\ref{eq:kp}) of the main text and the
eigenfunctions
\begin{equation*}
\begin{aligned}
\phi_{nk}(x)&= |A_{nk}|\left[e^{i\alpha_{nk}\left(x-\frac{a}{2}\right)}\right. \\
&\left.-\frac{1-e^{ia(\alpha_{nk}-k)}}{1-e^{-ia(\alpha_{nk}+k)}}
e^{-i\alpha_{nk}\left(x+\frac{a}{2}\right)} \right] \,,
\label{eq:KP-EFs}
\end{aligned}
\end{equation*}
with the normalization factor
\begin{equation*}
\left|A_{nk}\right|^2=\frac{\sin^2 \left[a (\alpha_{nk}
+k)/2\right]}{ 1-\frac{\sin (2 a \alpha_{nk} )}{2 a \alpha_{nk}
}-\cos (a k) \left[\cos (a \alpha_{nk} )-\frac{\sin (a
\alpha_{nk})}{a \alpha_{nk} }\right]}
\end{equation*}
and corresponding eigenenergies
\begin{equation}
\varepsilon_{nk}=\frac{\alpha_{nk}^2}{2} \,. \label{eq:kp_def_ener}
\end{equation}
For crystal momenta ${k=0}$ and even $n$, and for ${k=\pi/a}$ and
odd $n$, we have ${\alpha_{nk}=n\pi/a}$.

We adjust the global phase of the eigenfunctions to yield real
matrix elements in Eq.~(\ref{eq:pnnp}) of the main text,
\begin{equation*}
P^{nn'}_k=\frac{2|A_{nk}||A_{n'k}|\alpha_{nk}\alpha_{n'k}\left[ \cos
(a\alpha_{n'k})- \cos (a\alpha_{nk}) \right]
}{a\sin\left[\frac{a}{2}(\alpha_{nk}+k)\right]
\sin\left[\frac{a}{2}(\alpha_{n'k}+k)\right]\left(\alpha_{nk}^2-\alpha_{n'k}^2\right)}
\end{equation*}
for ${n\neq n'}$, and real diagonal matrix elements
\begin{equation*}
P^{nn}_k=\frac{\alpha_{nk}  \sin (a \alpha_{nk} ) \sin (a
k)}{1-\frac{\sin (2 a \alpha_{nk} )}{2 a \alpha_{nk} }-\cos (a k)
\left[\cos (a \alpha_{nk} )-\frac{\sin (a \alpha_{nk} )}{a
\alpha_{nk} }\right]} \,.
\end{equation*}

\section{\label{app_an_cont}Analytical continuation of the Kronig-Penney basis}

In our analysis of the interband current, we extend the energies in
Eq.~(\ref{eq:kp_def_ener}) to the complex ${K-}$plane, defining
${K=k+ik_i}$, where ${k_i}$ is the imaginary part of the complex
crystal momentum, and request
\begin{equation*}
\lim_{k_i\to 0}\alpha_{nK}=\alpha_{nk} \;\;\;, {\lim_{k_i\to
0}\varepsilon_{nK}=\varepsilon_{nk}} \,.
\end{equation*}
Equation~(\ref{eq:kp}) and its $K$-derivative at ${K=0}$yield
\begin{equation*}
\begin{aligned}
1&= \cos (a \alpha_{n0} )+ \frac{V_0}{\alpha_{n0}} \sin (a \alpha_{n0})\\
0&= \left[\frac{a\,V_0\cos (a
\alpha_{n0})}{\alpha_{n0}}-\left(a+\frac{V_0}{\alpha_{n0}^2}\right)\sin
(a \alpha_{n0})\right]\frac{\text{d}\alpha_{n0} }{\text{d} K}.
\end{aligned}
\end{equation*}
After eliminating  ${V_0}$ in these two equations, we see that the
term in square brackets of the second equation cannot be zero, since
${2\pi\leq a\alpha_{n0}<3\pi}$ for ${n=v\,,c}$. This implies
\begin{equation}
\left.\frac{\text{d}\alpha_{nK} }{\text{d}
K}\right|_{K=0}=0\;,\;\;\;\;n=v,\,c \label{eq:der_comp_alph}
\end{equation}
and shows that
\begin{equation}
\left.\frac{\text{d}\left(\Delta\varepsilon_{K}^{cv}\right)}{\text{d}K}\right|_{K=0}=0
\,. \label{eq:first_der_cond}
\end{equation}

From the second derivative of Eq.~(\ref{eq:kp}) with respect to $K$
and  Eq.~(\ref{eq:der_comp_alph}), we obtain
\begin{equation*}
-a^2= \left[ \frac{a\,V_0\cos (a
\alpha_{n0})}{\alpha_{n0}}-\left(a+\frac{V_0}{\alpha_{n0}^2}\right)\sin
(a \alpha_{n0})\right]\frac{\text{d}^2\alpha_{n0}}{\text{d} K^2} \,.
\end{equation*}
As noted in the main text,  the coefficients in front of the second
derivative are real. They do not vanish, since $2\pi\leq
a\alpha_{n0}<3\pi$ for ${n=v\,,c}$. From this we conclude that
\begin{equation*}
\Im  \left[\left.\frac{\text{d}^2\alpha_{nK} }{\text{d}
K^2}\right|_{K=0}\,\right]=0\;,\;\;\;\;n=v,\,c
\end{equation*}
and, consequently,
\begin{equation*}
\Im  \left[\left.\frac{\text{d}^2\left(\Delta
\varepsilon^{cv}_{K}\right)}{\text{d}
K^2}\right|_{K=0}\,\right]=0\,.
\end{equation*}

\section{\label{app:aux} Alternative expression for the interband current}
We here derive the interband current in Eq.~(\ref{eq:inter_two}) of
in the main text. Defining the function
\begin{equation}
\eta_k^{er}=\rho^{cv}_k(t) D^{vc}_{\kappa(t)}e^{-i  S(k,\,t)} \,,
\label{eq:etaer}
\end{equation}
we employ Eq.~(\ref{eq:tdm}) to obtain its derivative
\begin{equation*}
\begin{aligned}
\dot{\eta}_k^{er}&=\dot{\rho}^{cv}_k(t) D^{vc}_{\kappa(t)}e^{-i  S(k,\,t)}  \\
&+\rho^{cv}_k(t) \dot{D}^{vc}_{\kappa(t)}e^{-i  S(k,\,t)}  \\
&-P^{vc}_{\kappa(t)}\rho^{cv}_k(t)\,e^{-i  S(k,\,t)} \,.
\end{aligned}
\end{equation*}
Using
${\left(D^{vc}_{\kappa(t)}\right)^2=-\left|D^{vc}_{\kappa(t)}\right|^2}$
and substitution  of ${\dot{\rho}^{cv}_k(t)}$ from
Eq.~(\ref{eq:polarization}) results in
\begin{equation*}
\begin{aligned}
-\rho^{cv}_k(t) P^{vc}_{\kappa(t)}e^{-i  S(k,\,t)}=&\dot{\eta}_k^{er}-i
\Delta\rho_k^{vc}(t) E(t) \left|D^{vc}_{\kappa(t)}\right|^2\\
-&\rho^{cv}_k(t) \dot{D}^{vc}_{\kappa(t)}e^{-i  S(k,\,t)}  \,.
\end{aligned}
\end{equation*}
Since the second term in this equation is purely imaginary, after
replacing the ${\eta^{er}_k(t)}$ from Eq.~(\ref{eq:etaer}),
Eq.~(\ref{eq:inter_two_a})  can be written as
\begin{equation}
\begin{aligned}
J^{er}_k(t)=&\frac{d}{dt}\left[\rho^{cv}_k(t) D^{vc}_{\kappa(t)}e^{-i  S(k,\,t)}\right]  \\
&-\rho^{cv}_k(t) \dot{D}^{vc}_{\kappa(t)}e^{-i S(k,\,t)}+\text{c.c.}
\,.  \label{eq:inter_sbe}
\end{aligned}
\end{equation}

%---New App D
\section{\label{app:saddle_a}Saddle-point approximation for the $k$-channel interband yield}

\begin{figure}[h]
\centering{}\includegraphics[width=0.8\linewidth]{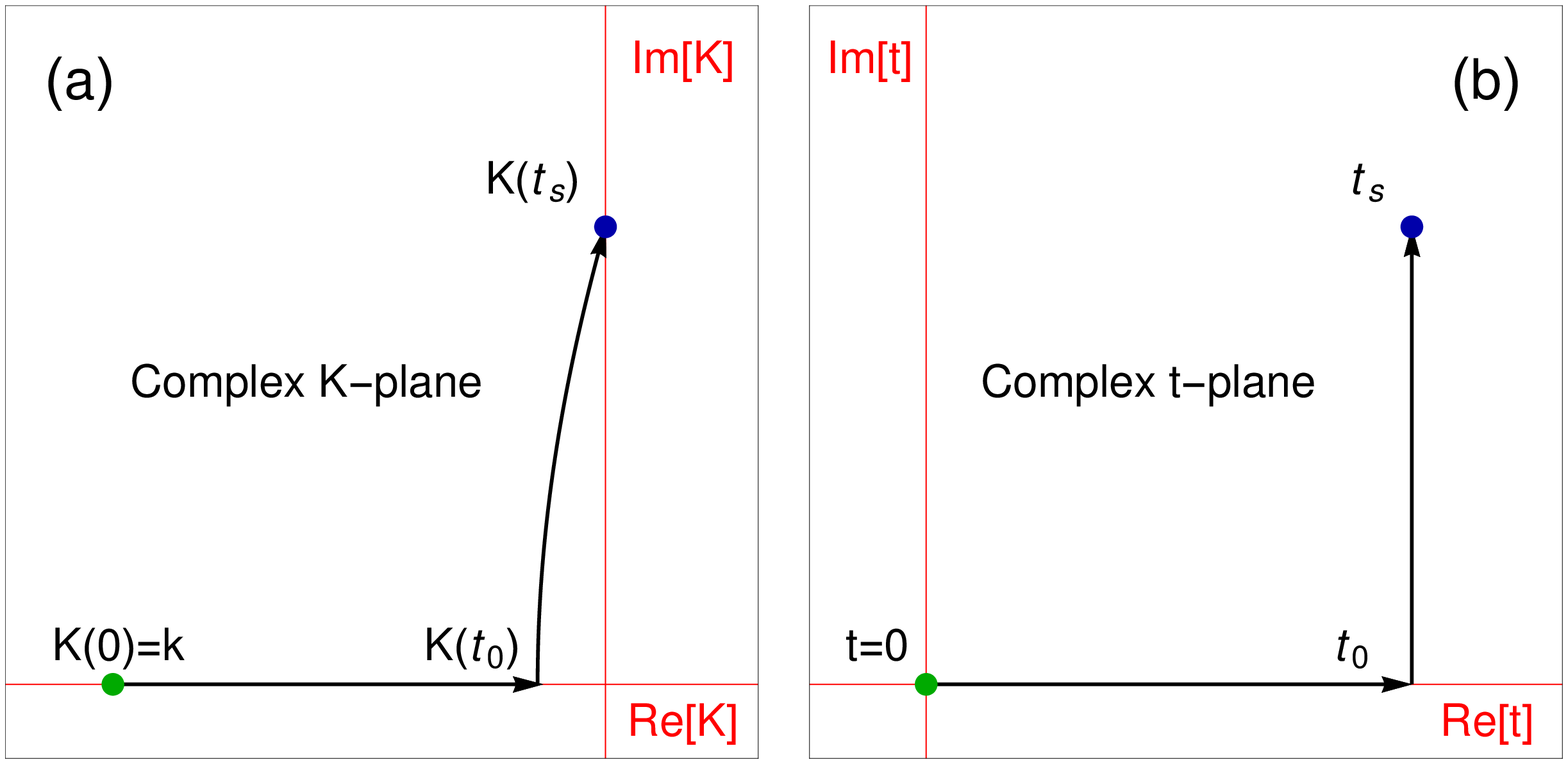}
\caption{Contour integration for the action evaluation.
(a) Integration path  in the complex crystal-momentum ($K-$)plane,
starting at ${K=k}$ (green dot), moving along the real $K-$axis to
${K(t_0)}$, and arriving at  the root of the analytic continuation
of the band-gap energy ${\Delta\varepsilon_{K}^{cv}}$, ${K(t_s)}$.
(b) Corresponding path in the complex temporal plane, starting at
time ${t=0}$ (green dot) and arriving at ${t_s}$ (blue dot).
 }
\label{fig:fig12}
\end{figure}

We evaluate the action in  Eq.~(\ref{eq:yield_a}) by splitting its
integral representation into two parts,
\begin{equation}
S(k,t_s)=\int_{0}^{t_0} \Delta\varepsilon_{K(t)}^{cv} dt
+\int_{t_0}^{t_s} \Delta\varepsilon_{K(t)}^{cv} dt\;,
\label{eq:int_im}
\end{equation}
with ${t_0=\Re [t_s]}$. Along the integration path illustrated in
Fig.~\ref{fig:fig12}, the first integral is real.

To evaluate the second integral, we need to calculate ${t_s}$. For
this purpose, we consider the continuum-wave vector potential
${A_0\sin(\omega_0 t)}$, for which we find
\begin{equation}
t_s=\frac{1}{\omega_0}\arcsin\left[-k/A_0 \pm i \gamma \right],
\label{eq:ts_a}
\end{equation}
with
\begin{equation*}
\gamma=\frac{\omega_0\sqrt{2\Delta\varepsilon_{0}^{cv} m^*_0}}{E_0}
\end{equation*}
and ${E_0=A_0\,\omega_0}$. For ${\gamma<1}$ and ${k<A_0<\pi/a}$,
Taylor expansion of Eq.~(\ref{eq:ts_a}) results in
\begin{equation}
\Im  [t_s]= \pm\frac{\gamma}{\omega_0\sqrt{1-(k/A_0)^2}}+
\mathcal{O}(\gamma ^3)\;, \label{eq:im_ts}
\end{equation}
which will be a fair approximation if ${\sqrt{1-(k/A_0)^2}<\gamma}$.

We can now carry out the integrations in Eq.~(\ref{eq:int_im}) and
obtain
\begin{equation}
\begin{aligned}
S(k,t_s)  &= \Re [S(k,\,t_s) ] +i \Delta\varepsilon_{0}^{cv}\Im  [t_s] \\
&+\frac{A^2}{8m^*_0}\left\{2\Im  [t_s]-\sinh\left\{2\omega_0 \Im  [t_s]\right\}/\omega_0\right\}\\
&+\frac{k^2}{2m^*_0}\left\{\Im  [t_s]-\tanh\left[ \omega_0 \Im
[t_s]\right\}/\omega_0\right\}  \label{eq:S-compex-plane} \,,
\end{aligned}
\end{equation}
where
 \begin{equation*}
\Re [S(k,t_s) ]=\int_{0}^{t_0} \Delta\varepsilon_{K(t)}^{cv} dt+\Re
\left[\int_{t_0}^{t_s} \Delta\varepsilon_{K(t)}^{cv} dt\right]\;.
\end{equation*}
To find the pre-exponential factor in Eq.~(\ref{eq:yield_a}), we
first point out that the cross derivatives ${\partial ^2
\sigma_\omega(k,\,t,\,t')/\partial t
\partial t'}$ and ${\partial ^2 \sigma_\omega(k,\,t,\,t')/\partial
t' \partial t}$ vanish, such that
\begin{equation*}
\begin{aligned}
\left|\text{det Hess}\right| &= \left|\frac{\partial ^2 \sigma_\omega(k,\,t_e,\,t_s)}{\partial t'\,^2}\frac{\partial ^2 \sigma_\omega(k,\,t_e,\,t_s)}{\partial t^2} \right| \\
&=\left| E(t_s)\sqrt{\frac{2\Delta\varepsilon_{0}^{cv}}{m^*_0}}
E(t_e)\left(P^{cc}_{\kappa(t_e)}-P^{vv}_{\kappa(t_e)}\right)\right|
\;.
\end{aligned}
\end{equation*}
Following~\cite{becker2002,lew1994pra}, only keeping  the regular
solution  in Eq.~(\ref{eq:im_ts}), ${\Im  [t_s]>0}$, we obtain
\begin{equation*}
\begin{aligned}
\hat{\bar{J}}^{er}_k (\omega)&\approx-\left(2\pi\omega \right)\left|D^{vc}_0\right|^2
e^{-\Im  [S(k,\,t_s)]}  \\
&\times \sum_{t_e} \frac{i\,E(t_s) e^{-i\omega t_e}e^{i\Re
[S(k,\,t_s)]}e^{i S(k,\,t_e)}}{\left|
E(t_s)\sqrt{\frac{2\Delta\varepsilon_{0}^{cv}}{m^*_0}}
E(t_e)\left(P^{cc}_{\kappa(t_e)}-P^{vv}_{\kappa(t_e)}\right)\right|^{1/2}}.
\end{aligned}
\end{equation*}
The interband HHG yield is thus given by
\begin{equation*}
\begin{aligned}
Y^{er}_k (\omega)&\approx\left(2\pi\omega \right)^2 e^{-2\Im
[S(k,\,t_s)]}\left|D^{vc}_0\right|^4 E(t_s)
\sqrt{\frac{m^*_0}{2\Delta\varepsilon_{0}^{cv}}} \\
&\times \left|\sum_{t_e} \frac{e^{-i\omega t_e}\left[ e^{i\left\{\Re
[S(k,\,t_s)]+
S(k,\,t_e)+\frac{\pi}{2}\right\}}+\text{c.c.}\right]}{\left|
E(t_e)\left(P^{cc}_{\kappa(t_e)}-P^{vv}_{\kappa(t_e)}\right)\right|^{1/2}}\right|^2
\;.
\end{aligned}
\end{equation*}

In order to get a more explicit equation in terms of the band and
laser field parameters, we approximate ${E(t_s)}$ to first order in
$\gamma$,
\begin{equation}
E(t_s)\approx E_0 \sqrt{1-(k/A_0)^2} +\mathcal{O}(\gamma^2)  \;,
\end{equation}
and apply the approximations used to derive ${\Im  [S(k,\,t_s)]}$ in
Eq.~(\ref{eq:S-compex-plane}). This leads to the interband yield
\begin{equation*}
\begin{aligned}
Y^{er}_k (\omega)&\approx
\exp\left[-\frac{2\,\sqrt{2 m^*_0}\,{\Delta\varepsilon_{0}^{cv}}^{3/2}}{E_0\sqrt{1-(k/A_0)^2}}\right]  \\
&\times\left(2\pi\omega \right)^2\left|D^{vc}_0\right|^4 E_0
\sqrt{1-(k/A_0)^2} \sqrt{\frac{m^*_0}{2\Delta\varepsilon_{0}^{cv}}} \\
&\times \left|\sum_{t_e} \frac{e^{-i\omega t_e}\left[ e^{i\left\{\Re
[S(k,\,t_s)]+
S(k,\,t_e)+\frac{\pi}{2}\right\}}+\text{c.c.}\right]}{\left|
E(t_e)\left(P^{cc}_{\kappa(t_e)}-P^{vv}_{\kappa(t_e)}\right)\right|^{1/2}}\right|^2
\;.
\end{aligned}
\end{equation*}

\section{\label{app:saddle_kr}   Simplification of Eq.~(\ref{S_c})}

We here approximate Eq.~(\ref{S_c}), proceeding in a similar way as
in Appendix \ref{app:saddle_a}, by splitting Eq.~(\ref{S_c}) in a
real and a complex integral (Fig.~\ref{fig:fig12})
\begin{equation*}
\int_{t_s}^{t_e}\Delta P^{vc}_{\kappa_r (t)}dt =I_1+I_2 \,,
\end{equation*}
with
\begin{equation*}
\begin{aligned}
I_1&=\int_{t_s}^{t_0}\Delta P^{vc}_{\kappa_r (t)} dt\\
I_2&= \int_{t_0}^{t_e}\Delta P^{vc}_{\kappa_r (t)} dt \,.
\end{aligned}
\end{equation*}
Following the derivation of Eq.~(\ref{eq:egK}), we find
\begin{equation*}
\frac{d\left(\varepsilon^{cv}_K\right)}{dK}\equiv - \Delta P^{vc}_K
=\frac{K}{m^*_0} + \mathcal{O}(K^3) \,.
\end{equation*}
Using ${\kappa(t)=k+A(t)}$, we  perform the integral in $I_1$ by
contour integration in the complex plane to obtain
\begin{equation*}
\begin{aligned}
I_1&=\frac{k}{i\omega_0 m_0^*}\left\{\tanh \left[\omega_0 \Im  (t_s)\right]-\omega_0\Im  (t_s)\right\} \\
&-\frac{A_0}{\omega_0 m_0^*}\cos(\omega_0 t_0)\left\{1-\cosh
\left[\omega_0 \Im  (t_s)\right]\right\} \\
&\approx
\mathcal{O}(\gamma^3)
 \;,
\end{aligned}
\end{equation*}
where we used that, according  Eq.~(\ref{eq:egK}), ${K(t_s)}$ is
imaginary. Thus, to second order in $\gamma$,  ${I_1\approx 0}$ and
Eq.~(\ref{S_c}) can be approximated as the real integral
\begin{equation*}
\int_{t_s}^{t_e}\Delta P^{vc}_{\kappa_r (t)} dt \approx
\int_{t_0}^{t_e}\Delta P^{vc}_{\kappa_r (t)} dt \,.
\end{equation*}
This expression  is more suitable for numerical calculations than
Eq.~(\ref{S_c}).

\section*{Acknowledgements}

This work was supported in part by the Air Force Office of
Scientific Research under award number ${{\text
{FA9550-17-1-0369}}}$,  NSF Grants ${{\text {No. 1464417}}}$ and
${{\text {No. 1802085}}}$, the ``High Field Initiative'' ${{\text
{(CZ.02.1.01/0.0/0.0/15\_003/0000449)}}}$ of European Regional
Development Fund (HIFI), and the project ``Advanced research using
high intensity laser produced photons and particles'' ${{\text
{(CZ.02.1.01/0.0/0.0/16\_019/0000789)}}}$ through the European
Regional Development Fund (ADONIS). Any opinions, finding, and
conclusions or recommendations expressed in this material are those
of the authors and do not necessarily reflect the views of the
United States Air Force.

%{\dkgreen{Underlines are introduced by the "ulem" package, which
%provides the "sout" (text crossing) function. When we remove that
%package for compilation in the final version, the underlining in
%references will be removed as well.}

%\bibliography{hhg_solids_withoutURL}% Produces the bibliography via BibTeX.
\providecommand{\noopsort}[1]{}\providecommand{\singleletter}[1]{#1}%

\end{document}